\documentclass[aps,prd,preprint,nofootinbib]{revtex4}
\usepackage{epsfig}
\usepackage{amsmath}
\usepackage{amssymb}
\usepackage{amsbsy}
\usepackage{amsfonts}
\usepackage{graphics}
\usepackage{latexsym}
\usepackage{dcolumn}
\usepackage{wasysym}
\usepackage{slashed}
\usepackage[dvips]{color}
\usepackage{graphicx}
\usepackage[figuresright]{rotating}

\begin{document}

\preprint{
\vbox{
\vspace*{0.5cm}
\hbox{DESY 06-114}
\hbox{Edinburgh 2006/07}
\hbox{LU-ITP 2006/012}
\hbox{LTH 706}
}}
\vspace*{1cm}

\title{Hadron spectrum, quark masses and decay constants from light overlap
  fermions on large lattices}

\author{D. Galletly$^1$, M. G\"urtler$^2$, R. Horsley$^1$, H. Perlt$^3$,
  P.E.L. Rakow$^4$, \\ 
G. Schierholz$^{2,5}$, A. Schiller$^3$ and T. Streuer$^{2,6}$}

\affiliation{\vspace*{0.3cm} $^1$ School of Physics, University of Edinburgh,
  Edinburgh EH9 3JZ, UK\\
$^2$ John von Neumann-Institut f\"ur Computing NIC, Deutsches
Elektronen-Synchrotron DESY, D-15738 Zeuthen, Germany\\
$^3$ Institut f\"ur Theoretische Physik, Universit\"at Leipzig, D-04109
Leipzig, Germany\\
$^4$ Theoretical Physics Division, Department of Mathematical
Sciences, University of Liverpool, Liverpool L69 3BX, UK\\
$^5$ Deutsches Elektronen-Synchrotron DESY, D-22603 Hamburg, Germany\\
$^6$Institut f\"ur Theoretische Physik, Freie Universit\"at Berlin,
D-14196 Berlin, Germany}

\author{- QCDSF--UKQCD Collaboration -}
\affiliation{ }

\begin{abstract} 
We present results from a simulation of quenched overlap fermions with
L\"uscher-Weisz gauge field action on lattices up to $24^3 \, 48$ and for
pion masses down to $\approx 250$ MeV. Among the quantities we study are the
pion, rho and nucleon masses, the light and strange quark masses, and the
pion decay constant. The renormalization of the scalar and axial vector
currents is done nonperturbatively in the $RI-MOM$ scheme. The simulations are
performed at two different lattice spacings, $a \approx 0.1$ fm and $\approx
0.15$ fm, and on two different physical volumes, to test the scaling
properties of our action and to study finite volume effects. We compare our
results with the predictions of chiral perturbation theory and compute several
of its low-energy constants. The pion mass is computed in sectors of fixed
topology as well. 
\end{abstract}


\maketitle

\section{Introduction}

Lattice simulations of QCD at small quark masses require a fermion action
with good chiral properties. Overlap fermions~\cite{Neuberger} possess an
exact chiral symmetry on the lattice~\cite{Luscher}, and thus are well suited
for this task. Furthermore, overlap fermions are automatically O(a) improved
if employed properly~\cite{QCDSF1}.

Previous calculations of hadron observables from quenched overlap fermions
have been limited to larger quark masses and/or coarser lattices due to the
high cost of the simulations~\cite{Rebbi,Liu,QCDSF2,Bietenholz}. To ensure
that the correlation functions this involves are not overshadowed by the 
exponential decay of the overlap operator~\cite{Hernandez}, the lattice
spacing $a$ should be small enough such that $m_H a \ll 2$ for mesons and $m_H
a \ll 3$ for baryons, where $m_H$ is the mass of the
hadron. In addition, the spatial extent of the lattice $L$ should satisfy 
$L \gg 1/(2 f_\pi)$ in order to be able to make contact with chiral
perturbation theory~\cite{Colangelo}. 

Over the past years we have done extensive simulations of quenched overlap
fermions~\cite{QCDSF2,QCDSF3,QCDSF4}. Furthermore, we have employed overlap
fermions to probe the topological structure of the QCD vacuum at
zero~\cite{Weinberg} and at finite temperature~\cite{Weinberg}.  
In this paper we shall give the technical details of our calculations and
present results on hadron and quark masses and the pseudoscalar decay constant,
including nonperturbative renormalization of the scalar, pseudoscalar and
axial vector currents. The bulk of the simulations are done on the $24^3 \,
48$ lattice at lattice spacing $a \approx 0.1$ fm. Our results on the spectral
properties of the overlap operator~\cite{QCDSF2} and nucleon structure
functions~\cite{QCDSF3} will be reported elsewhere in detail. 

The paper is organized as follows. In Section II we discuss the action and how
it is implemented numerically. In Section III we give the parameters of the
simulation. In Section IV we present our results for the hadron masses and the
pseudoscalar decay constant. The latter is used to set the scale. We compare
our results with the predictions of chiral perturbation theory, and attempt 
to compute some of its low-energy constants. In Section V we compute the
renormalization constants of the scalar and pseudoscalar density, as well as
the axial vector current, nonperturbatively, and in  Section VI we
present our results for the light and strange quark masses. Finally, in
Section VII we conclude.  

\section{The Action}

The massive overlap operator is defined by
\begin{equation}
D=\left(1-\frac{am_q}{2\rho}\right)\, D_N + m_q 
\label{D}
\end{equation}
with the Neuberger-Dirac operator $D_N$ given by
\begin{equation}
D_N=\frac{\rho}{a}\left(1+\frac{D_W(\rho)}{\sqrt{D_W^\dagger(\rho)
    D_W(\rho)}}\right)\,, \quad D_W(\rho)=D_W-\frac{\rho}{a}\,,
\label{DN}
\end{equation}
where $D_W$ is the massless Wilson-Dirac operator with $r=1$, and $\rho \in
[0,2]$ is a (negative) mass parameter. The operator $D_N$ has $n_- + n_+$
exact zero modes, $D_N \psi_n^0 = 0$ with $n=1, \cdots , n_- + n_+$,  
where $n_-$ ($n_+$) denotes the number of modes with negative (positive)
chirality, $\gamma_5 \psi_n^0 = - \psi_n^0$ ($\gamma_5 \psi_n^0 = +
\psi_n^0$). The index of $D_N$ is thus given by $\nu = n_- - n_+$. 
The `continuous' modes $\lambda_i$, $D_N \psi_i = \lambda_i \psi_i$,
satisfy $(\psi_i^\dagger,\gamma_5 \psi_i^{\phantom{\dagger}}) = 0$ and come in
complex conjugate pairs $\lambda_i^{\phantom{*}}, \lambda_i^*$. 

To evaluate $D_N$ it is appropriate to introduce the hermitean Wilson-Dirac
operator $H_W(\rho) = \gamma_5 D_W(\rho)$, such that
\begin{equation}
D_N = \frac{\rho}{a}\left(1+\gamma_5 \, {\rm sgn}\{H_W(\rho)\}\right)\,,
\end{equation}
where ${\rm sgn}\{H\} = H/\sqrt{H^2}$. The sign function can be defined by
means of the spectral decomposition
\begin{equation}
{\rm sgn}\{H_W(\rho)\} = \sum_i {\rm sgn}\{\mu_i\}\,
\chi_i^{\phantom{\dagger}} \chi^\dagger_i \,, 
\label{spec}
\end{equation}
where $\chi_i$ are the normalized eigenvectors of $H_W(\rho)$ with eigenvalue
$\mu_i$. Equation (\ref{spec}) is, however, not suitable for numerical
evaluation. We write
\begin{equation}
{\rm sgn}\{H_W(\rho)\} = \sum_{i=1}^N {\rm sgn}\{\mu_i\}\,
\chi_i^{\phantom{\dagger}} \chi^\dagger_i + P_\perp^N {\rm sgn}\{H_W(\rho)\}
\,, 
\end{equation}
where 
\begin{equation}
 P_\perp^N = 1 - \sum_{i=1}^N \chi_i^{\phantom{\dagger}}
 \chi^\dagger_i  
\end{equation}
projects onto the subspace orthogonal to the eigenvectors of the $N$ lowest
eigenvalues of $|H_W(\rho)|$, and 
approximate $P_\perp^N {\rm sgn}\{H_W(\rho)\}$ by a minmax
polynomial~\cite{Giusti}. More 
precisely, we construct a polynomial $P(x)$, such that 
\begin{equation}
\left|P(x)-\frac{1}{\sqrt{x}}\right| < \epsilon \, , \quad  x \in
[\mu^2_{N+1},\mu^2_{\rm max}] \,, 
\end{equation}
where $\mu_{N+1}$ ($\mu_{\rm max}$) is the lowest nonzero (largest)
eigenvalue of $|P_\perp^N H_W(\rho)|$. We then have 
\begin{equation}
{\rm sgn}\{H_W(\rho)\} = \sum_{i=1}^N {\rm sgn}\{\mu_i\}\,
\chi_i^{\phantom{\dagger}} \chi^\dagger_i + P_\perp^N H_W(\rho)\,
P(H_W^2(\rho))\,. 
\end{equation}
The degree of the polynomial will depend on $\epsilon$ and on the condition
number of $H_W^2(\rho)$, $\kappa = \mu^2_{\rm max}/\mu^2_{N+1}$, on the
subspace $\{\chi_i\,|\,(1- P_\perp^N) \chi_i = 0\}$.


We use the L\"uscher-Weisz gauge action~\cite{LW}
\begin{equation}
\begin{split}
S[U]&=\frac{6}{g^2}\left[ c_0 \sum_{\rm plaquette}\frac{1}{3}\,
\mbox{Re}\,\mbox{Tr}\, (1-U_{\rm plaquette}) + c_1 \sum_{\rm
  rectangle}\frac{1}{3}\, \mbox{Re}\, 
\mbox{Tr}\, (1-U_{\rm rectangle}) \right. \\
&+ c_2  \left. \sum_{\rm parallelogram}\frac{1}{3}\,
\mbox{Re}\, \mbox{Tr}\, (1-U_{\rm parallelogram}) \right] \,, 
\end{split}
\end{equation}
where $U_{\rm plaquette}$ is the standard plaquette, $U_{\rm rectangle}$
denotes the closed loop along the links of the $1 \times 2$ rectangle, and
$U_{\rm parallelogram}$ denotes the closed loop along the diagonally opposite
links of 
the cubes. The coefficients  $c_1$, $c_2$ are taken from tadpole improved
perturbation theory~\cite{Gattringer}:  
\begin{equation}
\frac{c_1}{c_0} = - \frac{(1+0.4805 \alpha)}{20 u_0^2} \,, \quad 
\frac{c_2}{c_0} = - \frac{0.03325 \alpha}{u_0^2}
\end{equation}
with $c_0 + 8 c_1 + 8 c_2 = 1$, where
\begin{equation}
u_0 = \left(\frac{1}{3} {\rm Tr}\, \langle U_{\rm plaquette} \rangle
\right)^{\frac{1}{4}} \,, \quad \alpha = - \frac{\log(u_0^4)}{3.06839} \,.
\end{equation}
We write
\begin{equation}
\beta = \frac{6}{g^2} \, c_0 \,.
\end{equation}
After having fixed $\beta$, the parameters $c_1$, $c_2$ are determined. In the
classical continuum limit $u_0 \rightarrow 1$ the coefficients $c_1$, $c_2$
assume the tree-level Symanzik values~\cite{Symanzik} $c_1=-1/12$, $c_2=0$.



\section{Simulation Parameters}

The simulations are done on the following lattices: 
\begin{equation}
\begin{tabular}{c|c|c} 
$\beta$ & Volume & $r_0/a$ \\ \hline 
$8.00$ & $16^3 \, 32$ & $3.69(4)$ \\ \hline 
$8.45$ & $16^3 \, 32$ & $5.29(7)$ \\ \hline 
$8.45$ & $24^3 \, 48$ & $5.29(7)$ \\ 
\end{tabular}
\label{lattices}
\end{equation}
The scale parameter $r_0/a$ was taken from~\cite{Gattringer}. The couplings
have been chosen such that the $16^3 \, 32$ lattice at $\beta=8.0$ and the
$24^3 \, 48$ lattice at $\beta=8.45$ have approximately the same physical
volume. This allows us to study both scaling violations and finite size
effects.  

We have
projected out $N=40$ lowest lying eigenvectors at $\beta = 8.0$ and $N=50$
($N=10$) at $\beta=8.45$ on the $24^3 \, 48$ ($16^3 \, 32$) lattice. These
numbers scale roughly with the physical volume of the lattice. The degree of
the polynomial $P$ has been adjusted such that $1/\sqrt{H_W^2(\rho)}$ is
determined with a relative accuracy of better than $10^{-7}$.

\begin{figure}[b]
\vspace*{-0.5cm}
\centering
  \epsfig{file=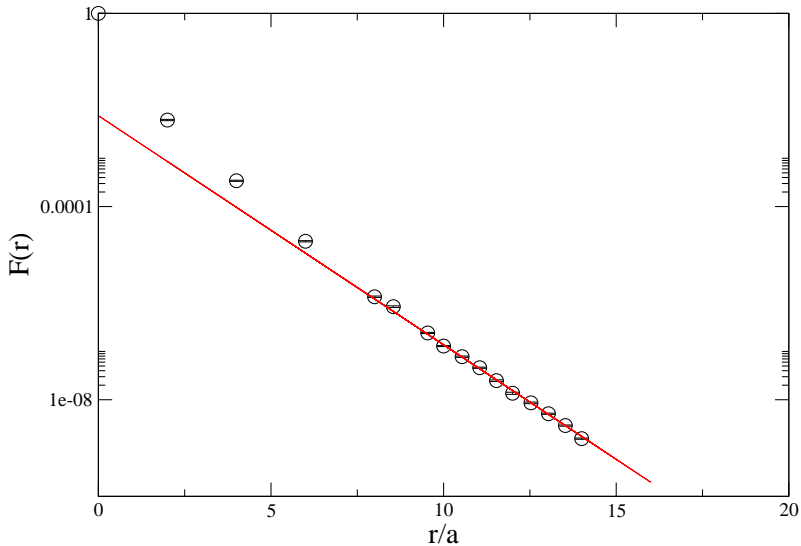,angle=270,width=11cm,clip=}
  \caption{The effective range $F(r)$ as a function of $r/a$ on the $16^3 \,
  32$ lattice at $\beta=8.45$ for $\rho=1.4$, together with an exponential
  fit. The fit gave $\mu = 1.11(1)$.}
  \label{fig:loc}
\end{figure}

The mass parameter $\rho$ influences the simulation in two ways. First, it
affects the locality properties~\cite{Hernandez} of the Neuberger-Dirac
operator. In Fig.~\ref{fig:loc} we show the effective range of $D_N$,  
\begin{equation}
F(r)= \left\langle \left\langle \max_x |D_N(x,y)|\,\Big|_{||x-y||=r}
    \right\rangle_y  \,\right\rangle_U \,,
\end{equation}
with respect to the Euclidean distance
\begin{equation}
||x|| = \left(\sum_{\mu=1}^4 x_\mu^2 \right)^{\frac{1}{2}} \,.
\end{equation}
\begin{figure}[b]
\vspace*{-2.3cm}
\centering
  \epsfig{file=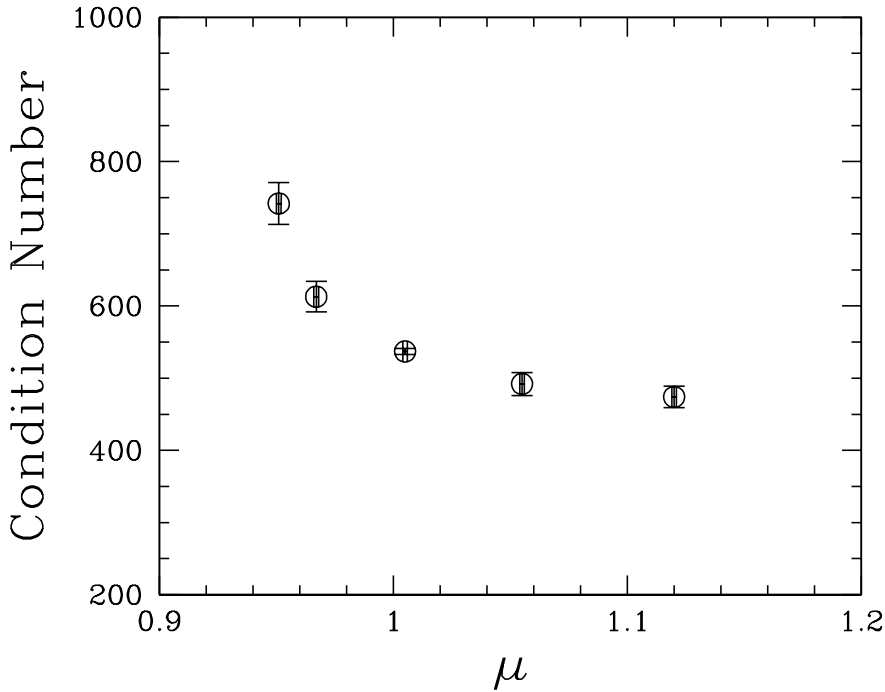,width=12cm,clip=}
  \caption{Condition number and $\mu$ on the $16^3 \, 32$ lattice at
    $\beta=8.45$ for $\rho=1.2$, $1.3$, $1.4$, $1.5$ and $1.6$, from left to
    right.} 
  \label{fig:cond}
\end{figure}
Asymptotically, $F(r) \propto \exp{-\mu r/a}$, where $\mu$ depends (among
others) on
$\rho$. (Numerically, $\mu \approx 2\, \nu$, where $\nu$ refers to the taxi
driver distance~\cite{Hernandez}.) We want $\mu$ to be as large as possible,
in particular $2 \mu \gg m_H a$ ($3 \mu \gg m_H a$) for mesons
(baryons). Secondly, the condition number of $P_\perp^N 
H_W^2(\rho)$, $\kappa = \mu^2_{\rm max}/\mu^2_{N+1}$, depends on $\rho$ as
well. In 
Fig.~\ref{fig:cond} we show the $\rho$ dependence of $\mu$ and $\kappa$ on the
$16^3 \, 32$ 
lattice at $\beta=8.45$ for $N=10$. Test runs show, however, that $\kappa$ does
not decrease significantly anymore if we increase $N$ further. We have chosen
$\rho = 1.4$, which is a trade-off between a small condition number $\kappa$
and a large value of $\mu$. At this value of $\rho$ we find $\mu = 1.11(1)$, 
which is consistent with the results obtained in~\cite{Hernandez} from the
Wilson gauge action.  

The simulations are performed at the following quark masses:
\begin{equation}
\vspace*{0.25cm}
\begin{tabular}{c|c|c|c|c|c|c|c|c} 
$\beta$ & $V$ & \multicolumn{7}{c}{$a m_q$} \\ \hline 
$8.00$ & $16^3\, 32$ & 0.0168 & 0.0280 & 0.0420 & 0.0560 & 0.0840 & 0.1400 &
0.1960 \\  
\hline
$8.45$ & $16^3\, 32$ & & \phantom{0.0112} & \phantom{0.0196} & 0.0280 & 0.0560
& 0.0980 & 0.1400 \\ 
$8.45$ & $24^3\, 48$ & & 0.0112 & 0.0196 & 0.0280 & 0.0560 & 0.0980 & 0.1400
\\  
\end{tabular}
\label{quarkmasses}
\end{equation}
This covers the range of pseudoscalar masses $250 \lesssim m_{PS} \lesssim 
900 \, \mbox{MeV}$ as we shall see. The lowest quark mass was chosen such that
$m_{PS} L \gtrsim 3$ ($L$ being the spatial extent of the lattice). On all our
lattices we have $L \gg 1/(2 f_\pi)$.

$O(a)$ improvement, both for masses and on- and off-shell operator matrix
elements, is achieved by simply replacing $D$ by~\cite{QCDSF1}
\begin{equation}
D^{\rm imp} \equiv \left(1-\frac{am}{2\rho}\right)\, D\,
  \left(1-\frac{a}{2\rho}  D\right)^{-1} 
\end{equation}
in the calculation of the quark propagator. Apart from the multiplicative mass
term, this amounts to subtracting the
contact term from the propagator. In the following we shall always
use the improved 
propagator, without mentioning it explicitly. The eigenvalues of $D_N$
lie on a circle of radius $\rho/a$ around $(\rho/a,0)$ in the complex plane,
while the eigenvalues of the improved operator $\displaystyle D_N^{\rm imp} =
D_N\left(1-\frac{a}{2\rho}D_N\right)^{-1}$ lie on the imaginary axis. 

\section{Hadron Masses and Pseudoscalar Decay Constant}

Let us now turn to the calculation of hadron masses and the pseudoscalar decay
constant. Before we can compare our results with the real world, we have to set
the scale. We will use the pion decay constant to do so, for reasons which will
become clear later. The pion decay constant derives from the axial vector
current, which has to be renormalized in the process. 
  
\subsection{Calculational Details}

The coefficients $c_1$, $c_2$ of the gauge field action are~\cite{Gattringer}
$c_1=-0.169805$, $c_2=-0.0163414$ at $\beta=8.0$ and $c_1=-0.154846$,
$c_2=-0.0134070$ at $\beta=8.45$. For the gauge field update we use a
heat bath algorithm, which we repeat 1000 times to generate a new
configuration. 

\clearpage
The inversion of the overlap operator $D$ is done by solving the system of
equations 
\begin{equation}
A x = y \, ,
\label{cg}
\end{equation}
where $A=D^\dagger D$ and $y$ is the relevant source vector. We use the
conjugate 
gradient algorithm for that. The speed of convergence depends on the condition
number of the operator $A$, $\kappa(A) = \nu_{\rm max}/\nu_{\rm min}$,
where $\nu_{\rm max}$ ($\nu_{\rm min}$) is the largest (lowest)
eigenvalue of $A$. For reasonable values of the quark mass we have $\kappa(A)
\propto 1/m_q^2$. Thus, the number of iterations, $n_D$, needed to achieve a
certain accuracy will grow like $n_D \propto 1/m_q$ as the quark mass is
decreased. 

The convergence of the algorithm can be accelerated by a preconditioning
method. Instead of (\ref{cg}) we solve the equivalent system of equations
\begin{equation}
A C x = C y \equiv \tilde{A} x\, ,
\label{pre}
\end{equation}
where $C$ is a nonsingular matrix, which we choose such that
$\kappa(\tilde{A}) \ll \kappa(A)$. Our choice is
\begin{equation}
C=1+\sum_{i=1}^{n} \left(\frac{1}{\nu_i} - 1 \right)
v_i^{\phantom{\dagger}} v_i^\dagger \, ,
\end{equation}
where $v_i$ ($\nu_i$) are the normalized eigenvectors (eigenvalues) of $A$.
The condition number of the operator $\tilde{A}$ is by a factor 
$\nu_{n+1}/\nu_1$ smaller than the condition number of the operator
$A$, and the number of iterations in the conjugate gradient algorithm
reduces to $n_D \propto 1/\sqrt{\nu_{n+1} + m^2_q}$, which depends only
weakly on the quark mass $m_q$. We have chosen $n=80$, and the inversion was
stopped when a relative accuracy of $10^{-7}$ was reached.

In the calculation of meson and baryon correlation functions we use smeared
sources to improve the overlap with the ground state, while the sinks are 
taken to be either smeared or local. We use Jacobi smearing for source and
sink~\cite{QCDSF5}. To set the size of the source, we have chosen
$\kappa_s=0.21$ for the smearing hopping parameter and employed $N_s=50$
smearing steps. 

To further improve the signal of the correlation functions, we have deployed
low-mode averaging~\cite{lma} in some cases by breaking the quark
propagator into two pieces, 
\begin{equation}
\sum_{i=1}^{n_\ell} \frac{\psi_i(x) \psi_i^\dagger (y)}{\lambda_i^{\rm imp} +
  m_q} \, ,  
\label{low}
\end{equation}
where the sum extends over the eigenmodes of the $n_\ell$ lowest eigenvalues
of $D_N^{\rm imp}$, and the remainder. The contribution from the 
low-lying modes (\ref{low}) is averaged over all positions of the quark
sources. As the largest contribution to the correlation functions comes from
the lower modes, we may expect a significant improvement in the regime of
small quark masses. We have chosen $n_\ell=40$, mainly because of memory
limitations. 

\begin{figure}[b]
\vspace*{1cm}
\centering
  \epsfig{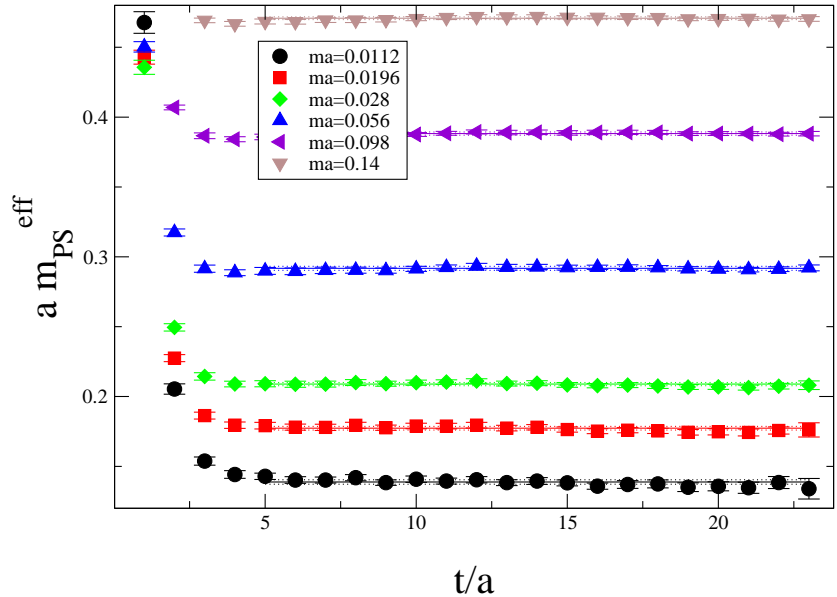}
  \caption{The effective pseudoscalar mass from the correlation function of
    the axial vector current $A_4$ on the $24^3\, 48$ lattice at $\beta=8.45$,
    using smeared sources and local sinks.} 
  \label{fig:mpieff}
\end{figure}

\subsection{Lattice Results}

The calculations are based on $900-1300$ gauge field configurations at the
lowest four quark masses at $\beta=8.0$, and on $200-300$ configurations
elsewhere. We consider hadrons only with all quarks having degenerate masses.

{\it Pion Mass}

To compute the pseudoscalar mass, $m_{PS}$, we looked at correlation functions
of the pseudoscalar density $P=\bar{\psi}\gamma_5 \psi$ and the time
component of the axial vector current $A_4=\bar{\psi}\gamma_4\gamma_5 \psi$. 
In Fig.\ref{fig:mpieff} we show the corresponding effective mass
for our four lowest quark masses on the $24^3\, 48$ lattice. Local sinks are
found to give slightly smaller error bars than smeared sinks, so that we will
restrict ourselves to this case. Both correlators give consistent results. We
will use the results from the axial vector current correlator here, because it
results in a wider plateau as the pseudoscalar correlator, in particular at 
the larger quark masses. 
We fit the correlator by the function $A\cosh\left(m_{PS}(t-T/2)\right)$,
where $T$ is the temporal extent of the lattice, over the region of the  
plateau. The results of the fit are listed in Table~\ref{tab:bare_res}.

\begin{table}[t]
\centering
\begin{tabular}{c|c|l|l|l|l|l||c}
$\beta$ & $V$ & \multicolumn{1}{c|}{$a m_q$} &\multicolumn{1}{c|}{$a m_{PS}$}
& \multicolumn{1}{c|}{$a m_V$} & \multicolumn{1}{c|}{$a m_N$} &
\multicolumn{1}{c||}{$a f_{PS}$} & \multicolumn{1}{c}{$m_{PS}$ [MeV]} \\ 
\hline
& & 0.0168 & 0.190(1) & 0.643(5)& 0.793(5)& 0.075(1) & 239(1)\\
& & 0.0280 & 0.235(1) & 0.64935)& 0.821(4)& 0.076(1) & 295(1)\\
& & 0.0420 & 0.281(1) & 0.65923) & 0.863(3)& 0.078(1) & 353(1)\\
$8.00$\, & \,$16^3 \, 32$\, & 0.0560 & 0.321(1) & 0.669(2) & 0.890(3) &
0.080(1) & 403(1)\\
& & 0.0840 & 0.388(1) & 0.695(3) & 0.952(7) & 0.082(1) & 488(1)\\
& & 0.1400 & 0.502(1) & 0.751(2) & 1.074(7) & 0.090(1) & 631(1)\\
& & 0.1960 & 0.599(1) & 0.815(1) & 1.188(7) & 0.097(1) & 753(1)\\
\hline
& & 0.0280 & 0.212(3) & 0.441(6)$^\ast$ & 0.595(6)$^\ast$ & 0.053(1) & 396(8)\\
& & 0.0560 & 0.289(2) & 0.482(4)$^\ast$ & 0.675(4)$^\ast$ & 0.058(1) & 545(4)\\[-2.5ex]
$8.45$\, & \,$16^3 \, 32$\, & & & & & &\\[-2.5ex]
& & 0.0980 & 0.384(2) & 0.537(4) & 0.784(7) & 0.064(1) & 727(4)\\
& & 0.1400 & 0.467(2) & 0.595(3) & 0.886(6) & 0.070(1) & 883(4)\\
\hline
& & 0.0112 & 0.139(1) & 0.429(6)$^\ast$ & 0.551(12)$^\ast$ & 0.051(1) & 264(4)\\
& & 0.0196 & 0.177(1) & 0.442(6)$^\ast$ & 0.572(11)$^\ast$ & 0.052(1)  & 336(2)\\
& & 0.0280 & 0.209(1) & 0.452(3)$^\ast$ & 0.600(10)$^\ast$ & 0.054(1)  & 396(2)\\[-2.5ex]
$8.45$\, & \,$24^3 \, 48$\, & & & & & &\\[-2.5ex]
& & 0.0560 & 0.292(1) & 0.481(3) & 0.674(12) & 0.058(1) & 551(2)\\
& & 0.0980 & 0.388(1) & 0.538(2) & 0.788(11) & 0.065(1) & 731(2)\\
& & 0.1400 & 0.412(1) & 0.597(1) & 0.892(11) & 0.071(1) & 887(2)\\
\end{tabular}
\vspace*{0.5cm}
\caption{Hadron masses and pseudoscalar decay constant. The numbers marked by
$\ast$ are obtained with low-mode averaging. To convert $m_{PS}$ to physical 
units, we have used the result in (\ref{fpiparms}). The error on $m_{PS}$
in the last column is purely statistical.}
\label{tab:bare_res}
\vspace*{0.5cm}
\end{table}

{\it Rho and Nucleon Mass}

To compute the vector meson mass, $m_V$, we explored correlation functions
of operators $V_i=\bar{\psi}\gamma_i \psi$ and $V^4_i=\bar{\psi}\gamma_i
\gamma_4\psi$ ($i=1,2,3$). We found that the operator $V_i$, in combination\\[-0.75em]


\clearpage
\begin{figure}[t]
\vspace*{0.5cm}
\centering
  \epsfig{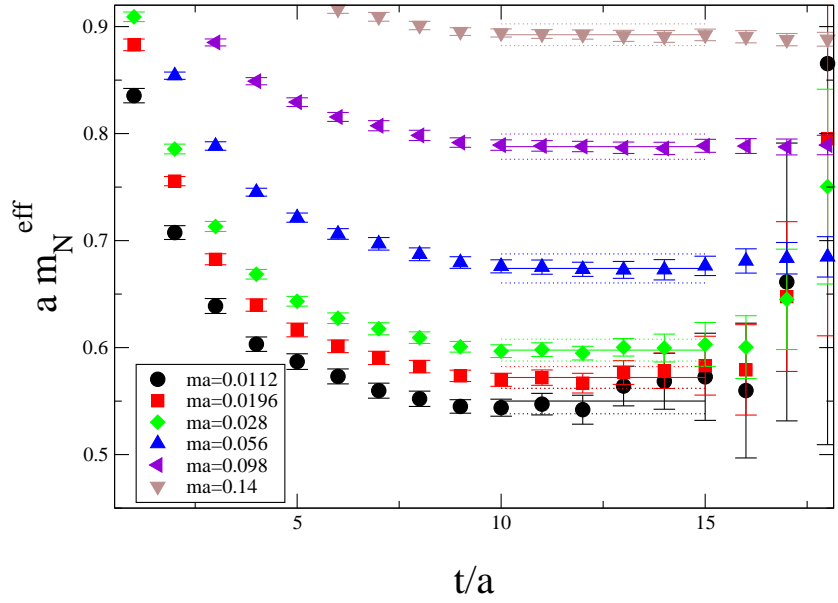}
  \caption{The effective nucleon mass on the $24^3\, 48$ lattice at
    $\beta=8.45$, using smeared sources and local sinks. The horizontal lines
    indicate the fit interval as well as the value and error of the mass. The
    data points at the lowest three quark masses have been computed with
    low-mode averaging.} 
  \label{fig:mneff}
\end{figure}
\begin{figure}[h]
\vspace*{1cm}
  \epsfig{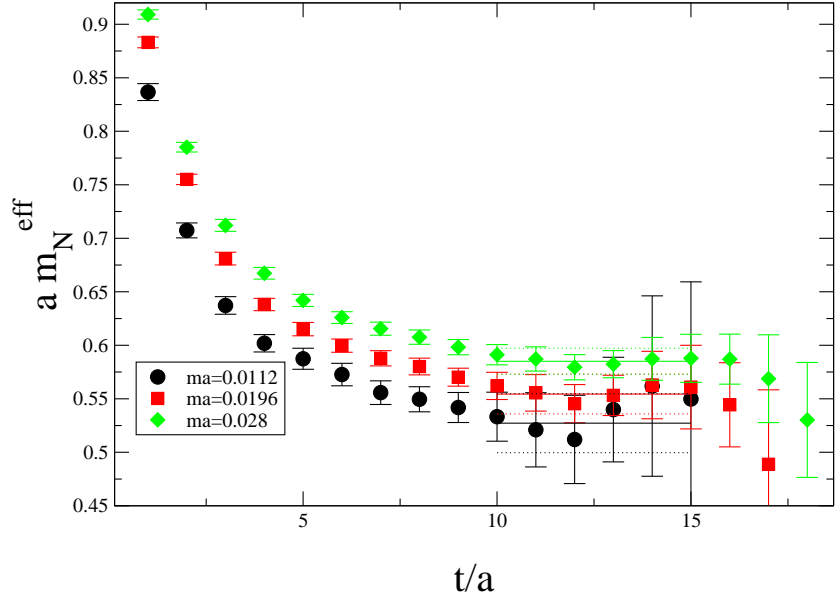}
  \caption{The same as the previous figure, but for the lowest three quark
  masses without using low-mode averaging.}  
  \label{fig:mneffsm}
\vspace*{0.3cm}
\end{figure}

\clearpage
\noindent
with a local sink, gives the best signal.

\begin{figure}[b]
\vspace*{-2.3cm}
\centering
  \epsfig{file=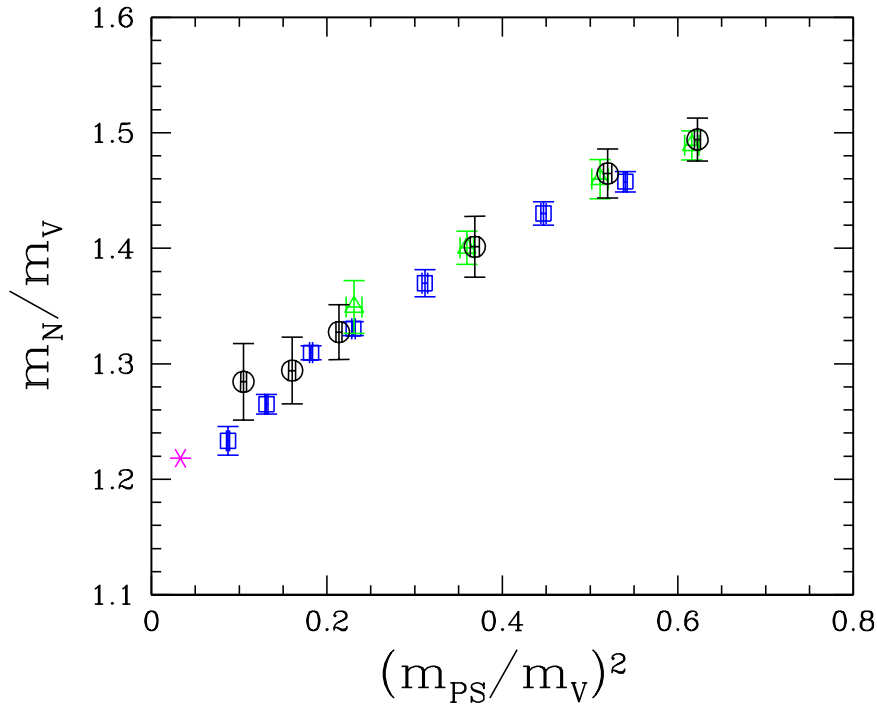,width=12cm,clip=}
  \caption{APE plot on the $24^3 \, 48$ lattice at $\beta=8.45$
    ($\Circle$) and on the $16^3 \, 32$ lattices at $\beta=8.0$
    (\textcolor{blue}{$\square$}) and 
    $\beta=8.45$ (\textcolor{green}{$\vartriangle$}).}  
  \label{fig:ape}
\vspace*{0.3cm}
\end{figure}

For the calculation of the nucleon mass, $m_N$, we used $B_\mu =
\varepsilon_{abc} \psi_\mu^a (\psi^b C\gamma_5 \psi^c)$ (where
$C=\gamma_4\gamma_2$) 
as our basic operator, where we have replaced each spinor by $\psi \rightarrow
\psi^{NR} = (1/2) (1+\gamma_4) \psi$~\cite{QCDSF5}. These so-called
nonrelativistic wave functions have a better overlap with the ground state
than the ordinary, relativistic ones. In Fig.~\ref{fig:mneff} we show the
effective nucleon mass for all our six quark masses on the $24^3\, 48$
lattice, where for the lowest three quark masses we have employed low-mode
averaging. We find good to reasonable plateaus starting at $t/a \gtrsim 8$. 
In Fig.~\ref{fig:mneffsm} we show, for comparison, the result obtained without
low-mode averaging. In this case the situation is less favorable. The nucleon
mass is obtained from a fit of the data by the correlation function $A
\exp(-m_Nt) + B \exp(-m_{N^*}(T-t))$, where $m_{N^*}$ is the mass of the
backward moving baryon, over the region of the plateau.

The results for the rho and nucleon masses are listed in
Table~\ref{tab:bare_res}. Note that $a m_V \ll 2 \mu$ and $a m_N \ll 3 \mu$,
respectively, are satisfied in all 
cases. In Fig.~\ref{fig:ape} we show an APE plot for our three lattices. At
our smallest quark masses we have $m_{PS}/m_V \approx 0.3$. The APE plot shows
no scaling violations outside the error bars and no finite size effects. 

{\it Pion decay constant}

The physical pion decay constant is given by
\begin{equation}
\langle 0|\mathcal{A}_4|\pi \rangle = m_\pi f_\pi\, ,
\end{equation}
where $\mathcal{A}_\mu$ is the renormalized axial vector current,
$\mathcal{A}_\mu = Z_A A_\mu$. Using the axial Ward identity
\begin{equation}
\partial_\mu \mathcal{A}_\mu = 2 m_q P\, ,
\end{equation}
where $P$ is the {\em local} pseudoscalar density, and considering the fact
that  $m_q P$ is a renormalization group invariant, we obtain
\begin{equation}
f_{\pi} = \frac{2 m_q}{m^2_{\pi}} \langle 0|P|\pi \rangle \, .
\end{equation}
On the lattice we consider the correlation function 
\begin{equation}
\begin{split}
\langle P^s(t) P^{s^\prime}(0) \rangle &= \frac{1}{2 a m_{PS}} \langle
0|P^s|PS\rangle 
\langle PS|P^{s^\prime}|0\rangle\, \left[\exp(-m_{PS}\, t) + \exp(-m_{PS}\,
  (T-t))\right] \\[0.4em] 
&\equiv A^{ss^\prime}\, \left[\exp(-m_{PS}\, t) + \exp(-m_{PS}\, (T-t))
  \right] \, ,
\end{split}
\end{equation}
where the superscripts $s,s^\prime$ distinguish between local ($L$) 
and smeared ($S$) operators. From this we obtain
\begin{equation}
a f_{PS} = a m_q \left(\frac{2}{am_\pi}\right)^{3/2}
\frac{A^{LS}}{\sqrt{A^{SS}}} \, . 
\end{equation}
We thus find $a f_{PS}$ by computing $A^{LS}$ and $A^{SS}$. In
Table~\ref{tab:bare_res} we give our results. In our notation the experimental
value of $f_\pi$ is $92.4$ MeV.

Comparing our data on the $16^3 \, 32$ and $24^3 \, 48$ lattice at
$\beta=8.45$ in Table~\ref{tab:bare_res} piece by piece, we also find no 
finite size effects down to the lowest common pseudoscalar mass.

\subsection{Setting the Scale: Pion Decay Constant}

\begin{figure}[t]
\vspace*{-2.3cm}
\centering
  \epsfig{file=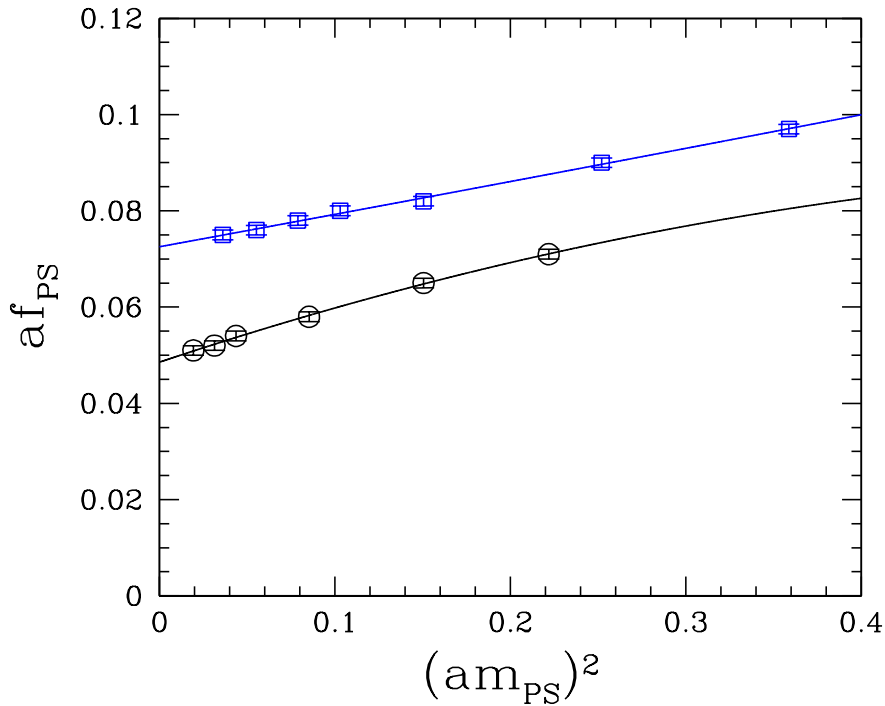,width=12cm,clip=}
  \caption{Chiral extrapolation of the pseudoscalar decay constant on the
  $24^3 \, 48$ lattice at $\beta=8.45$ 
    ($\Circle$) and on the $16^3 \, 32$ lattice at $\beta=8.0$
    (\textcolor{blue}{$\square$}).}   
  \label{fig:fpi}
\vspace*{0.3cm}
\end{figure}

We will use the pseudoscalar decay constant to set the scale. The reason is
that $f_{PS}$ is an analytic function in $m_{PS}^2$ for degenerate quark
masses~\cite{Pallante}, in contrast to $m_V$ and $m_N$, which exhibit
nonanalytic behavior. We thus expect that $f_{PS}$ extrapolates smoothly to
the chiral limit. In quenched chiral perturbation theory~\cite{Bernard,Sharpe}
to NLO we have\footnote{Here and in the following we shall adopt the
  notation $\alpha_i^q = 128 \pi^2 L_i^q$, $L_i^q$ being the conventional
  Gasser-Leutwyler coefficients~\cite{Gasser}. The superscript $q$ stands for quenched.}~\cite{Pallante}    
\begin{equation}
f_{PS} = f_0 \left(1+ \alpha_5^q \frac{m_{PS}^2}{2(4\pi f_0)^2} \right)
+O(m_{PS}^4) \, .
\end{equation}
In Fig.~\ref{fig:fpi} we show our data together with a quartic fit in the
pseudoscalar mass. The lattice spacing is obtained from requiring
$f_{PS} = f_\pi = 92.4$ MeV at the physical pion mass. Using the $r_0/a$
values given in (\ref{lattices}), we can convert the lattice spacing $a$
into the dimensionful scale parameter $r_0$. Altogether, we obtain
\begin{equation}
\begin{tabular}{c|c|c|c|c|c|c} 
$\beta$ & $V$ & $af_0$ & $\alpha_5^q$ & $a$ [fm] & $f_0$ [MeV] & $r_0$ [fm]\\
\hline  
$8.00$ & $16^3 \, 32$ & $0.073(1)$ & $1.5(4)$ & $0.157(3)$ & $92(1)$ &
$0.58(2)$\\ \hline  
$8.45$ & $24^3 \, 48$ & $0.049(1)$ & $1.9(4)$ & $0.105(2)$ & $91(2)$ &
$0.56(2)$\\  
\end{tabular}
\label{fpiparms}
\end{equation}
Note that $\alpha_5^q$, $f_0$ and $r_0$ come out independent of the lattice
spacing 
within the error bars, which, once more, indicates good scaling properties of
our action. The coefficient $\alpha_5^q$ turns out to be in
agreement with the phenomenological value of $1.83$ ($L_5^q=0.00145$)
reported in~\cite{Bijnens}. 
 
\subsection{Comparison with Chiral Perturbation Theory}
\label{comcpt}

We shall now compare our results for the pseudoscalar, vector meson and
nucleon mass with the predictions of chiral perturbation theory and attempt to
extrapolate the lattice numbers to the chiral limit.

\begin{figure}[t]
\vspace*{-2.3cm}
\centering
  \epsfig{file=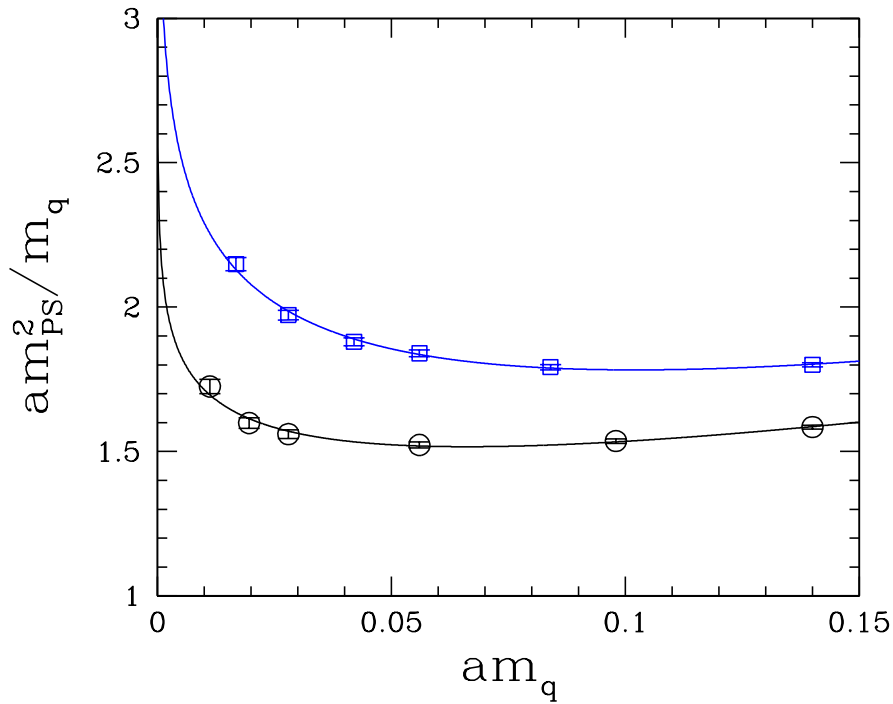,width=12cm,clip=}
  \caption{Chiral extrapolation of the pseudoscalar mass on the
  $24^3 \, 48$ lattice at $\beta=8.45$ 
    ($\Circle$) and on the $16^3 \, 32$ lattice at $\beta=8.0$
    (\textcolor{blue}{$\square$}). The curves show the fits for
    $\alpha_{\Phi}^q=0$.}    
  \label{fig:mpi}
\vspace*{0.3cm}
\end{figure}
 
{\it Pion Mass}

We plot the pseudoscalar masses as a function of the quark mass in
Fig.~\ref{fig:mpi}. Quenched chiral perturbation theory~\cite{Bernard}
predicts in the infinite volume~\cite{Pallante,Sommer} 
\begin{equation}
\frac{m_{PS}^2}{m_q} =A \left\{1-\left(\delta - \frac{2}{3}\alpha_{\Phi}^q\, y
  \right) \big( \ln\,y+1 \big)+\left[(2\alpha_8^q-\alpha_5^q) -
  \frac{\alpha_{\Phi}^q}{3} \right]y \right\}  + \cdots
\label{chipi}
\end{equation}
with 
\begin{equation}
y=\frac{A m_q}{\Lambda_{\chi}^2}
\end{equation}
where $A = 2 \Sigma/f^2_0$, $\Sigma$ being the `bare quark condensate', and  
$\Lambda_{\chi}$ denotes the scale at which the $\alpha_i$'s are being
evaluated. The traditional value is $\Lambda_{\chi} = 4\pi f_0$, which we will
also adopt here. For the parameter $\delta$ chiral perturbation theory
predicts~\cite{Bernard,Sharpe}
\begin{equation}
\delta = \frac{\mu_0^2}{48 \pi^2 f_{\pi}^2} \, ,
\end{equation}
with $\mu_0^2 \equiv m_{\eta^\prime}^2+m_{\eta}^2-2 m_K^2 = (870 \,
\mbox{MeV})^2$. This gives $\delta = 0.183$. The parameters $f_0$ and
$\alpha_5^q$ are known from our fit of $f_{PS}$ and are given in
(\ref{fpiparms}). 

\begin{figure}[t]
\vspace*{-2.3cm}
\centering
  \epsfig{file=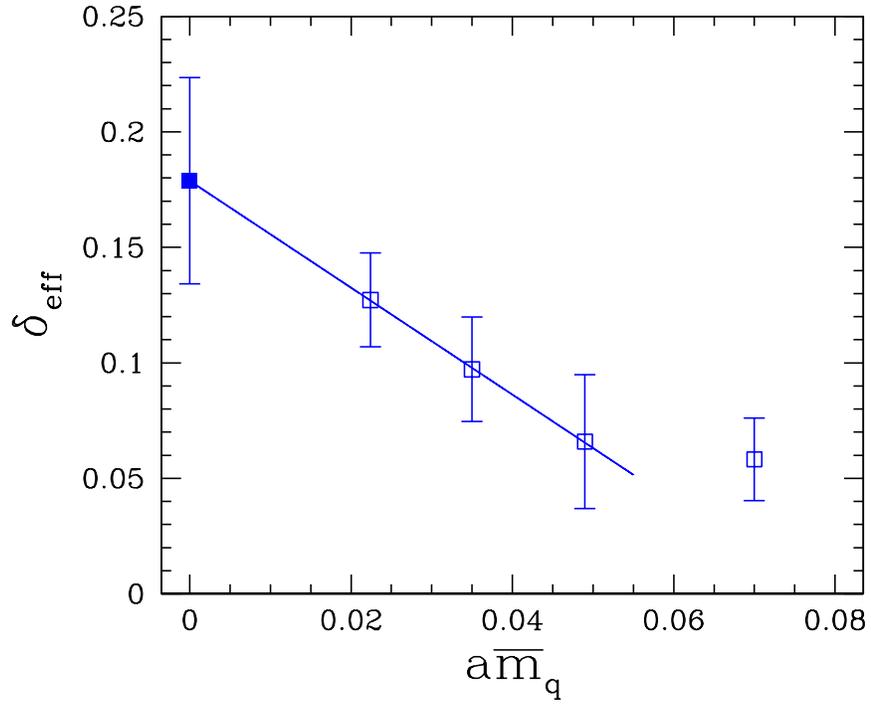,width=12cm,clip=}\vspace*{-1.5cm}
  \epsfig{file=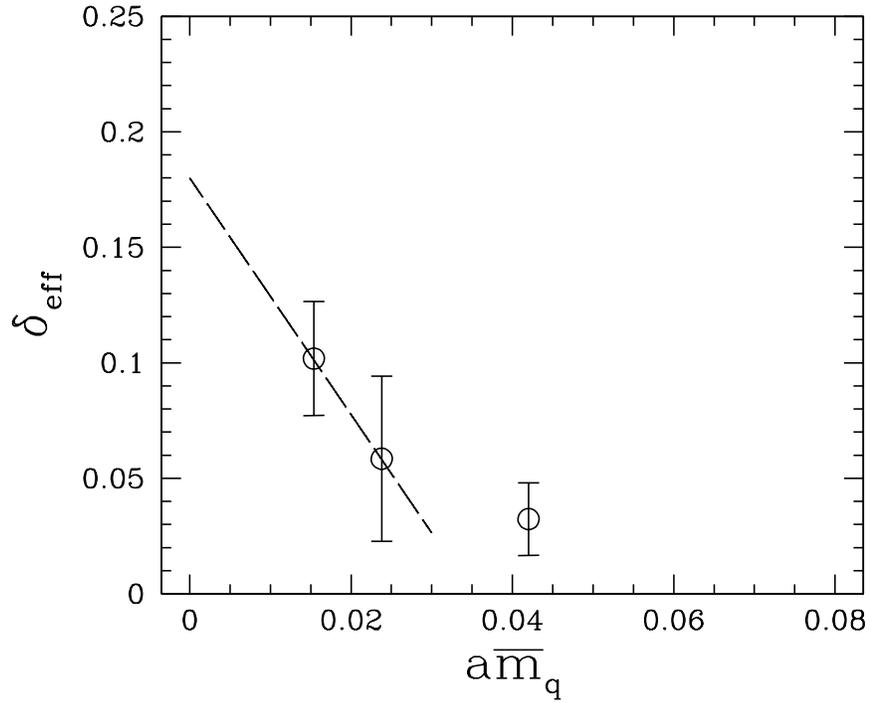,width=12cm,clip=}
  \caption{The parameter $\delta_{\rm eff}$ on the $16^3 \, 32$ lattice at
    $\beta=8.0$ together with a linear fit (top) and on the $24^3 \, 48$
    lattice at $\beta=8.45$ (bottom), as a function of the average quark mass
    $\bar{m}_q = (m_q + m_q^\prime)/2$.}
  \label{fig:mpid}
\vspace*{0.3cm}
\end{figure}

A much sought after quantity is the parameter $\delta$. Though unphysical, 
it would
be a great success of the calculation, and of quenched chiral perturbation
theory as well, if $\delta$ turned out to be in agreement with the predicted
value. We shall try to determine $\delta$ directly from the data. Let us write 
\begin{equation}
z=\frac{m_{PS}^2}{\Lambda_{\chi}^2} \, , \quad w=\frac{m_{PS}^2}{m_q} 
\end{equation}
and introduce the effective $\delta$ parameter
\begin{equation}
\delta_{\rm eff}^{-1} = 1 + \frac{\ln z^\prime \, w - \ln z \, w^\prime}{w -
  w^\prime} \, ,
\end{equation}
where $z, z^\prime$ and $w, w^\prime$, respectively, are adjacent data points.
It is easy to see that
\begin{equation}
\lim_{m_q \rightarrow 0} \delta_{\rm eff} = \delta \, .
\end{equation}
In Fig.~\ref{fig:mpid} we show $\delta_{\rm eff}$ as a function of the quark
mass. In the case of our high statistics run on the $16^3 \, 32$ lattice at
$\beta=8.0$ we are able to extrapolate $\delta_{\rm eff}$ to the chiral
limit. We obtain $\delta = 0.18(4)$, in agreement with the prediction of
quenched chiral perturbation theory. On the  $24^3 \, 48$ lattice at $\beta =
8.45$ our current statistics does not allow such an extrapolation. But the
data for $\delta_{\rm eff}$ are not inconsistent with the predicted value of
$\delta$.

The Witten-Veneziano formula~\cite{WV} relates $\mu_0^2$
to the topological susceptibility 
\begin{equation}
\chi_t= \frac{\langle Q^2\rangle}{V} \, ,
\end{equation}

\clearpage

\begin{figure}[t]
\vspace*{-2.3cm}
\centering
  \epsfig{file=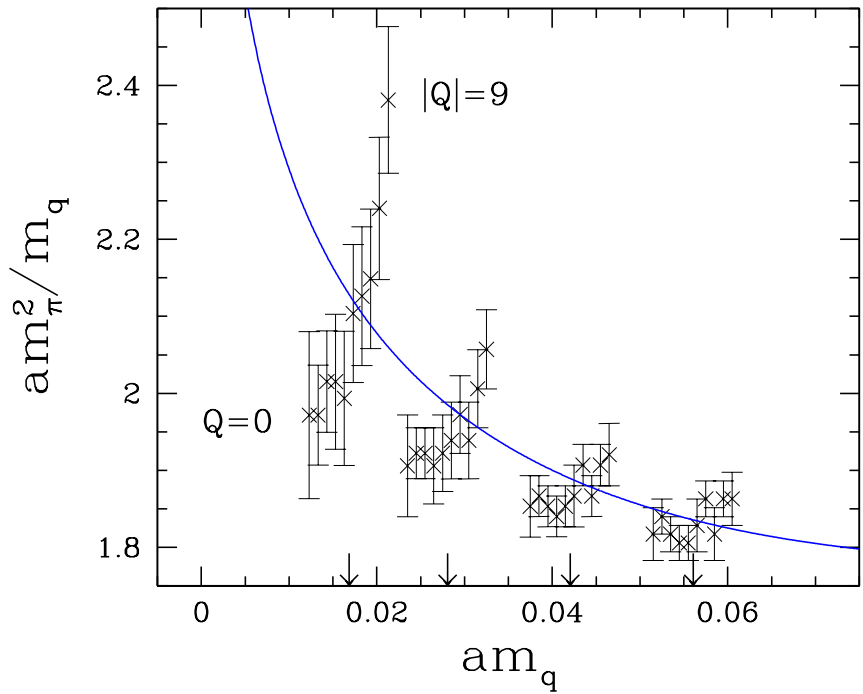,width=12cm,clip=}
  \caption{The pseudoscalar mass on the $16^3 \, 32$ lattice at $\beta=8.0$ for
    $|Q|=0, \cdots, 9$, from left to right. The data have been displaced
    horizontally. The true quark masses are indicated by the arrow at the
    bottom rim of the figure. The curve is from Fig.~\ref{fig:mpi}.} 
  \label{fig:mpiQ}
\vspace*{0.3cm}
\end{figure}

\clearpage
\noindent
where $Q$ is the topological charge and $V$ the lattice volume. The result for
$\delta$ is
\begin{equation}
\delta = \frac{1}{8 \pi^2 f_\pi^4} \, \chi_t \, , 
\end{equation}
which suggests that the pseudoscalar mass depends on the topological charge
$|Q|$. This turns out to be indeed the case. In Fig.~\ref{fig:mpiQ} we show the
pseudoscalar mass for various charge sectors, where the charge $Q$ is given by
the index $\nu$ of $D_N$. We observe a strong increase of
$\delta$ with increasing $|Q|$, and contrary to the findings in~\cite{Brower},
we do not expect the effect to go away in the limit $V \rightarrow \infty$,
$\chi_t$ fixed. It would
be interesting to search other quantities for a $|Q|$-dependence as well.

Let us now turn to the fit of (\ref{chipi}) to the data. Knowing $f_0$ and
$\alpha_5^q$, this leaves us with four free parameters. Because our
data do not allow an uncorrelated fit of all four parameters, we have to make
a choice and fix one of them. We consider two cases. In the first
case we fix $\alpha_{\Phi}$ at $0$, while in the second case we fix $\delta$ at
its theoretical value of $0.183$. The two fits give
\begin{equation}
\vspace*{0.25cm}
\begin{tabular}{c|c|c|c|c|c|c} 
$\beta$ & $V$ & $aA$ & $\delta$ & $\alpha_{\Phi}^q$ & $\alpha_8^q$ &
$\chi^2/{\rm dof}$\\ \hline  
 &  & $1.0(1)\phantom{0}$ & $0.34(7)$ & $\phantom{-}{\it 0.0}\phantom{(0)}$ &
 $1.4(6)$\ &  $0.8$\\[-1em]
 $8.00$ & $16^3 \, 32$ & & & & & \\[-1em]
 &  & $1.46(2)$ &  ${\it 0.183}\phantom{()}$ & $\phantom{-}0.8(2)$ &
 $1.5(2)$\ &  $1.3$\\ \hline
 &  & $1.22(4)$ & $0.16(2)$ & $\phantom{-}{\it 0.0}\phantom{(0)}$ & $1.3(2)$ &
 $1.0$  \\[-1em]  
$8.45$ & $24^3 \, 48$ & & & & & \\[-1em] 
 &  & $1.15(2)$ & ${\it 0.183}\phantom{()}$ & $-0.2(1)$ & $1.4(4)$ & $0.8$ \\ 
\end{tabular}
\label{mpifit}
\vspace*{0.25cm}
\end{equation}
where we have omitted the heaviest mass point at $\beta=8.00$. 
The numbers shown in {\it italics} are the numbers that we fixed. It is not
expected that $A$ scales. Assuming $\delta = 0.183$ and taking $f_0$ from
(\ref{fpiparms}), we obtain $a^3 \Sigma = 0.0039(1)$ at $\beta=8.0$ and $a^3
\Sigma = 0.00138(5)$ at $\beta=8.45$, respectively. We shall return to 
$\Sigma$ and the fit function (\ref{chipi}) when we compute the renormalized
chiral condensate and quark masses. Combining the results on both lattices, we
obtain $a_8^q = 1.5(4)$ for $\delta=0.183$. This is to be compared
with~\cite{Sommer} $a_8 = 0.8(4)$ in full QCD. In Fig.~\ref{fig:mpi} we
compare the fits with the data.

{\it Rho Mass}

\begin{figure}[t]
\vspace*{-2.3cm}
\centering
  \epsfig{file=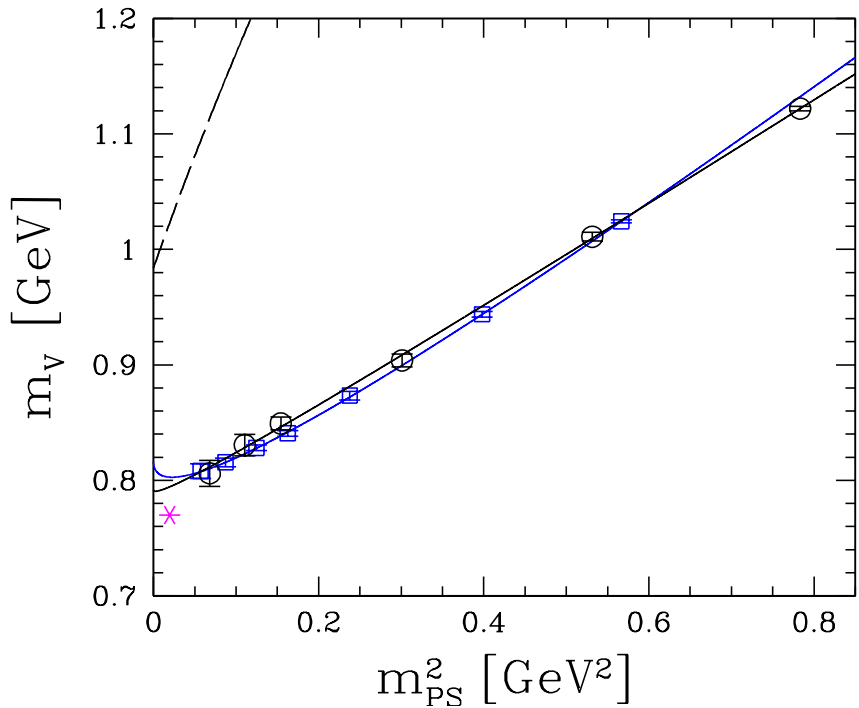,width=12cm,clip=}\\
  \caption{Chiral extrapolation of the vector meson mass on the
  $24^3 \, 48$ lattice at $\beta=8.45$ ($\Circle$) and on the $16^3 \, 32$
  lattice at $\beta=8.0$ (\textcolor{blue}{$\square$}), together with the
  experimental value (\textcolor{magenta}{$\ast$}). The solid curves show the
  fits. The dashed curve in the top left corner shows the energy of the state
  of two pseudoscalar mesons.} 
\label{fig:mrho}
\end{figure}

In Fig.~\ref{fig:mrho} we plot the vector meson masses as a function of the
pseudoscalar mass, where we have used the results of (\ref{fpiparms}) to
convert the lattice numbers to physical values. Quenched chiral perturbation
theory predicts~\cite{Booth} 
\begin{equation}
m_V = C_0^V + C_{1/2}^V\, m_{PS} + C_1^V\, m_{PS}^2 +
C_{3/2}^V\, m_{PS}^3 
+ \cdots \, , 
\label{chirho}
\end{equation}  
where $m_{PS}$ is the lattice pseudoscalar mass as described by (\ref{chipi}). 
The coefficient $C_{1/2}^V$ is expected to be negative, so that the
chiral limit is approached from below. Our data show no indication of a cubic
term, and so we shall drop that. A quadratic fit in the pseudoscalar 
mass gives
\begin{equation}
\vspace*{0.25cm} 
\begin{tabular}{c|c|c|c|c} 
$\beta$ & $V$ & $C_0^V \, [\mbox{GeV}]$ & $C_{1/2}^V$ & $C_1^V \,
[\mbox{GeV}^{-1}]$ \\ \hline  
$8.00$ & $16^3 \, 32$ &  $0.82(1)$ & $-0.18(5)$ & $0.61(5)$  \\
\hline   
$8.45$ & $24^3 \, 48$ & $0.79(2)$ & $-0.05(7)$ & $0.48(6)$ \\   
\end{tabular}
\label{rhoparms}
\vspace*{0.25cm}
\end{equation}
Our high statistics run at $\beta=8.0$ gives indeed a negative value for
$C_{1/2}$, but perhaps of lower magnitude than expected~\cite{Booth}, while at
$\beta=8.45$ our statistics is not high enough to make any statement. The fits
are shown in Fig.~\ref{fig:mrho}. One might think that at the lighter quark
masses one  is seeing the lowest two-pion state instead of the rho. In
Fig.~\ref{fig:mrho} we also show the energy of two pseudoscalar mesons at
the lowest nonvanishing lattice momentum\footnote{Note that the pions in the
rho are in a relative $p$ wave.}, $|p|=2\pi/(aL)$, assuming the lattice
dispersion relation to hold. We see that the lowest two-pion energy lies well
above the vector meson mass because of the finite size of our lattice.  

{\it Nucleon Mass}

\begin{figure}[t]
\vspace*{-2.3cm}
\centering
  \epsfig{file=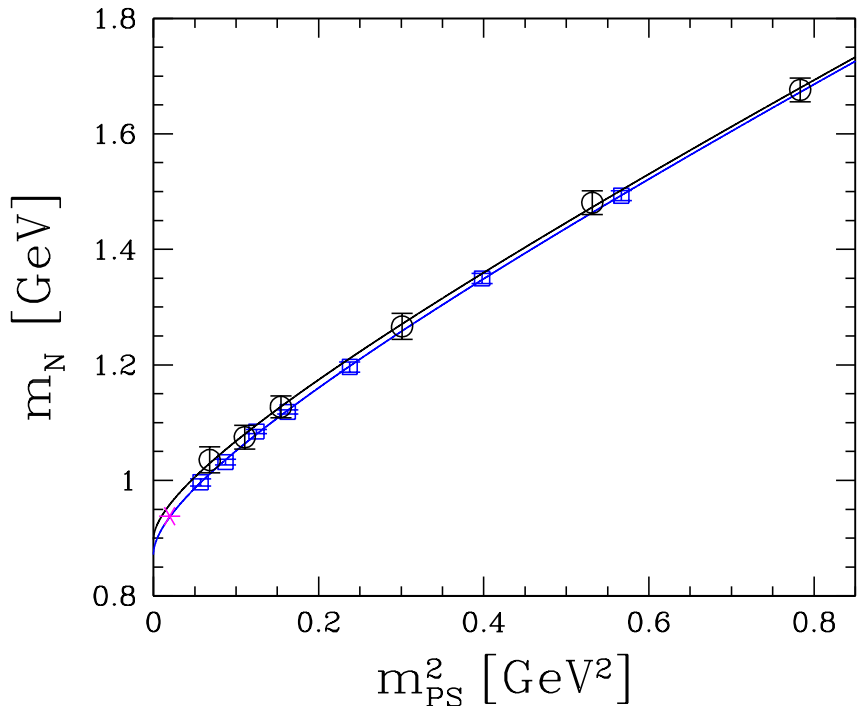,width=12cm,clip=}\\
  \caption{Chiral extrapolation of the nucleon mass on the
  $24^3 \, 48$ lattice at $\beta=8.45$ ($\Circle$) and on the $16^3 \, 32$
  lattice at $\beta=8.0$ ($\square$), together with the experimental value
  ($\ast$). The curves show the fits.}   
\label{fig:mn}
\end{figure}

We plot the nucleon masses as a function of the pseudoscalar mass in
Fig.~\ref{fig:mn}. Quenched chiral perturbation theory predicts~\cite{Labrenz}
\begin{equation}
m_N = C_0^N + C_{1/2}^N\, m_{PS} + C_1^N\, m_{PS}^2 + C_{3/2}^N\, m_{PS}^3
+ \cdots \, , 
\label{chimn}
\end{equation}  
where 
\begin{equation}
C_{1/2}^N = - \frac{3}{2} \left( 3F-D\right)^2 \pi \delta \, .
\end{equation}
Assuming the tree-level values $F=0.50$ and $D=0.76$, we expect $C_{1/2}^N = -
2.58 \,\delta$. For the theoretical value $\delta = 0.183$ this would give
$C_{1/2}^N = - 0.47$. Of course, $F$ and $D$ may be different in the quenched
theory. In the $N_c \rightarrow \infty$ limit, for example, $F/D = 1/3$ giving
$C_{1/2}^N = 0$. Again, our data show no indication of a cubic term, and we
shall 
drop that here as well. A quadratic fit in the pseudoscalar mass gives
\begin{equation}
\vspace*{0.25cm}
\begin{tabular}{c|c|c|c|c} 
$\beta$ & $V$ & $C_0^N \, [\mbox{GeV}]$ & $C_{1/2}^N$ & $C_1^N \,
[\mbox{GeV}^{-1}]$ \\ \hline  
$8.00$ & $16^3 \, 32$ & $0.87(2)$ & $0.4(1)$ & $0.6(1)$ \\
\hline   
$8.45$ & $24^3 \, 48$ & $0.90(7)$ & $0.3(3)$ & $0.6(2)$ \\   
\end{tabular}
\label{nucparms}
\vspace*{0.25cm}
\end{equation}
At $\beta=8.0$ we find some evidence for nonanalytic behavior, but with
positive coefficient $C_{1/2}^N$. The fits are shown in Fig.~\ref{fig:mn}.


Both, the vector meson and nucleon masses scale, within the error bars, with
the inverse lattice spacing set by the pion decay constant $f_\pi$.

\section{Nonperturbative Renormalization}

We shall now turn to the determination of the renormalization constants $Z_S$, 
$Z_P$ and $Z_A$ of the scalar and pseudoscalar density and the axial vector
current, respectively, which we will need in order to compute the renormalized
quark mass. We shall employ the $RI^{\prime}-MOM$ scheme~\cite{Martinelli}. 
Our implementation of this method is described in~\cite{QCDSF7}. 

\begin{figure}[b]
\vspace*{0.5cm}
  \centering
  \epsfig{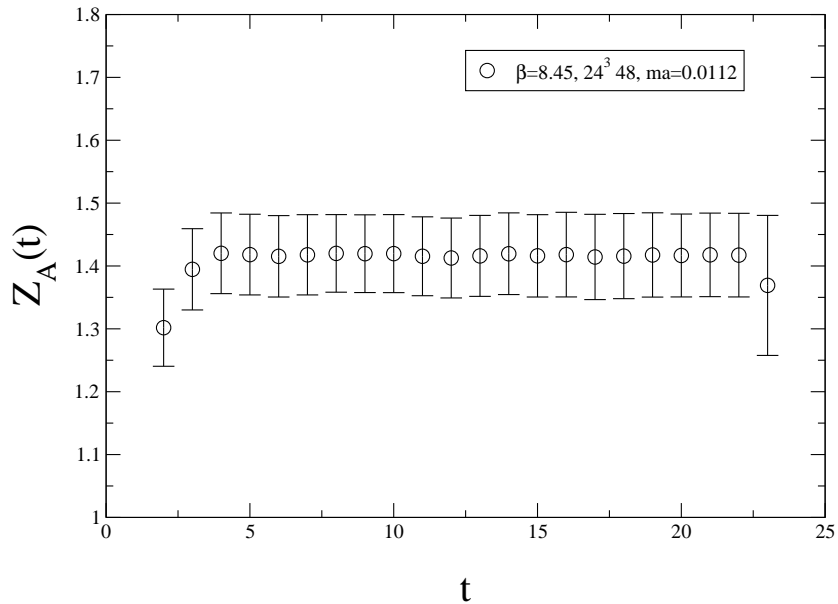}
  \caption{The renormalization constant $Z_A$ on the $24^3 \, 48$ lattice at
  $\beta=8.45$ at the smallest quark mass.}
  \label{fig:Z_A}
\end{figure}

We consider amputated Green functions, or vertex functions,
$\Gamma_{\mathcal{O}}$, with operator insertion $\mathcal{O} = S,\, P$ and
$A_4$ in the Landau gauge. Defining renormalized vertex functions by
\begin{equation}
\Gamma_{\mathcal{O}}^R(p) = Z_q(\mu)^{-1}
Z_{\mathcal{O}}(\mu)\, \Gamma_{\mathcal{O}}(p) \, ,
\end{equation}
where $\mu$ is the renormalization scale,
we fix the renormalization constants by imposing the renormalization condition 
\begin{equation}
\frac{1}{12}\, {\rm Tr} \left[\Gamma_{\mathcal{O}}^R(p) \,
  \Gamma_{\mathcal{O},\,{\rm Born}}^{-1}\right]\,\Big|_{p^2=\mu^2} = 1 \, .
\end{equation}
That is, we compute the renormalization constants from
\begin{equation}
Z_q(\mu)^{-1} Z_{\mathcal{O}}(\mu)\, \frac{1}{12}\, {\rm Tr}
\left[\Gamma_{\mathcal{O}}(p) \, 
\Gamma_{\mathcal{O},\,{\rm Born}}^{-1}\right]\Big|_{p^2=\mu^2} \equiv
Z_q(\mu)^{-1} Z_{\mathcal{O}}(\mu)\, 
\Lambda_{\mathcal{O}}(p)\Big|_{p^2=\mu^2}       
  = 1 
\end{equation}
with $\Gamma_{S,\,{\rm Born}} = 1$, $\Gamma_{P,\,{\rm Born}} = \gamma_5$ and
$\Gamma_{A,\,{\rm Born}} = \gamma_4 \gamma_5$. 

\begin{figure}[t]
\vspace*{-2.3cm}
  \centering
  \epsfig{file=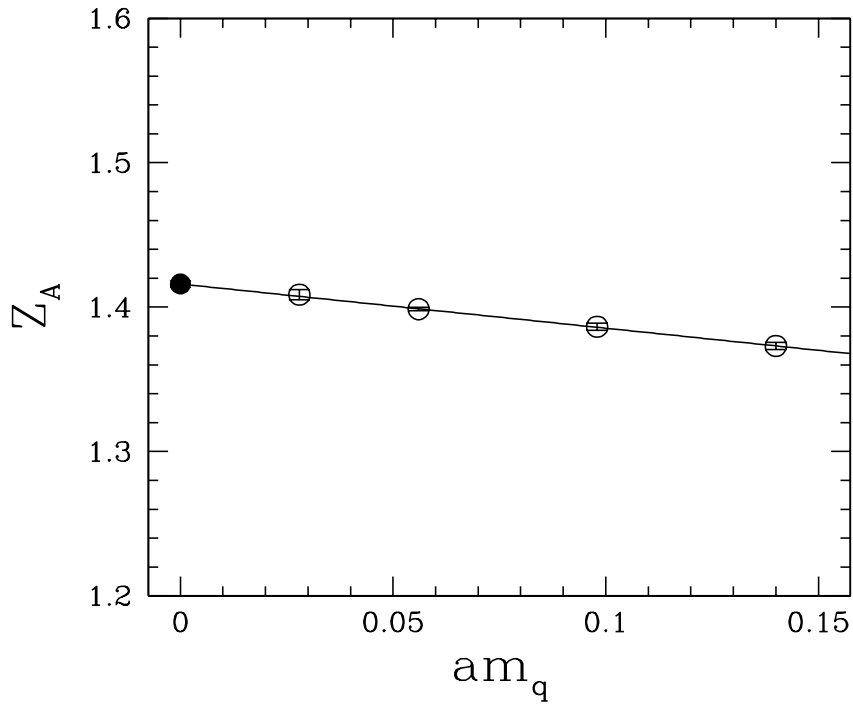,width=12cm, clip=}
  \caption{Chiral extrapolation of $Z_A$ on the $24^3 \, 48$ lattice at
  $\beta=8.45$, together with its value in the chiral limit ($\CIRCLE$) .}
  \label{fig:Z_Am}
\end{figure}

The renormalization constant of the axial vector current can be directly
determined from the axial Ward identity    
\begin{equation}
  Z_A = \frac{2 m_q \left<P\left(t\right)P\left(0\right)\right>}
  {\left<\partial_4 A_4\left(t\right)P\left(0\right)\right>} \, .
\label{Z_A}
\end{equation}
The wave function renormalization constant $Z_q$ can thus be obtained from
$\Lambda_A$ and $Z_A$,
\begin{equation}
Z_q(\mu) = Z_A \, \Lambda_A \, .
\end{equation}
In Fig.~\ref{fig:Z_A} we plot $Z_A$. We find that the r.h.s.\ of (\ref{Z_A}) is
independent of $t$, except for the points close to source and sink, as
expected. We extrapolate $Z_A$ linearly in $am_q$ to the chiral limit, as
shown in Fig.~\ref{fig:Z_Am}. The final result is
\begin{equation}
\begin{tabular}{c|c|c}
$\beta$ & $V$ & $Z_A$ \\ \hline
 $8.0\phantom{0}$&  $16^3\, 32$ & 1.59(1) \\
 $8.45$&  $16^3\, 32$ & 1.42(2) \\
 $8.45$&  $24^3\, 48$ & 1.42(1) \\
\end{tabular}
\label{tab:ZA}
\end{equation}
The corresponding fully tadpole improved (FTI) perturbative
numbers~\cite{QCDSF8} are $Z_A = 1.358$ at $\beta=8.0$ and $Z_A=1.303$ at
$\beta=8.45$. They lie $15\% - 8\%$ below their nonperturbative values. 

\begin{figure}[t]
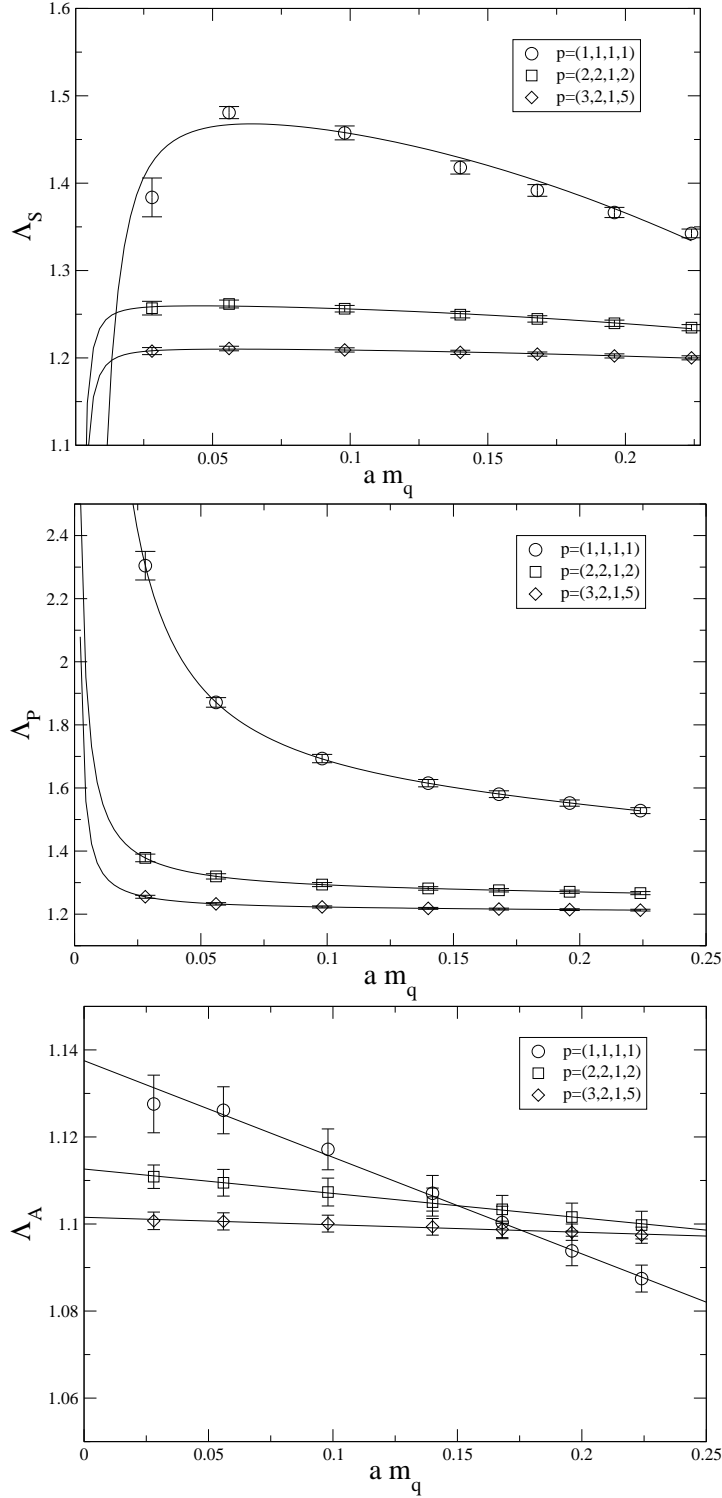

\vspace*{0.5cm}
  \centering
  \hspace*{-0.25cm}\epsfig{file=lambdas.eps,width=9.2cm,clip=}\\
  \epsfig{file=lambdap.eps,width=9.5cm,clip=}\\
  \epsfig{file=LambdaAV_gerrit.eps,width=9.5cm,clip=}
  \caption{Chiral extrapolation of $\Lambda_S$ (top), $\Lambda_P$ (middle) and
  $\Lambda_A$ (bottom) on the $16^3 \, 32$ lattice at $\beta=8.45$ for
  some representative momenta $p=(n_1,n_2,n_3,n_4)$ in units of $2\pi/aL$
  ($n_1, n_2, n_3$) and $\pi/aL$ ($n_4$).}
  \label{fig:Lambda_m}
\end{figure}

\begin{figure}[t]
\vspace*{0.5cm}
  \centering
  \epsfig{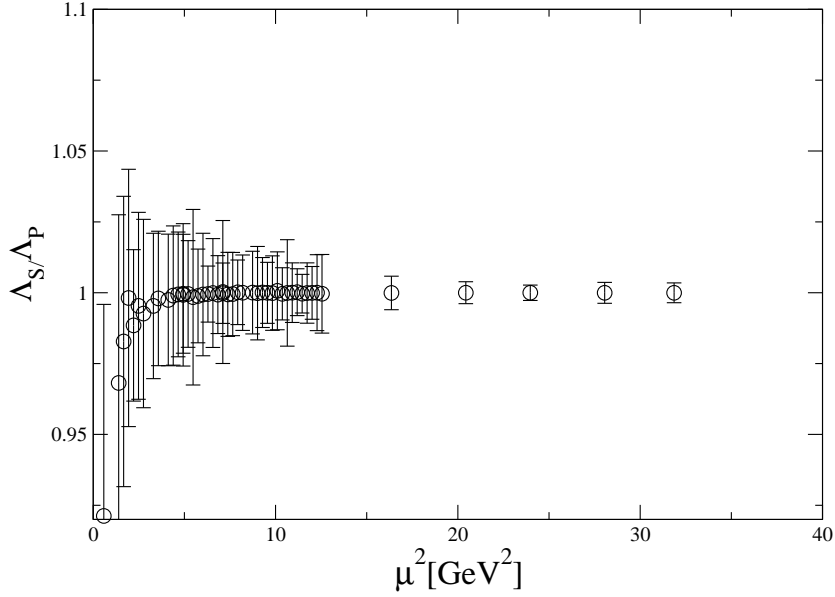}
  \caption{The ratio $\Lambda_S(\mu)/\Lambda_P(\mu)$ as a function of $\mu$ on
  the $16^3 \, 32$ lattice at $\beta=8.45$.}
  \label{fig:Lambda_S/P}
\end{figure}

Let us now turn to the calculation of $\Lambda_S(p)$, $\Lambda_P(p)$ and
$\Lambda_A(p)$. We denote the expressions at finite $m_q$ by
$\Lambda(p,m_q)$. Strictly speaking, $\Lambda_S(p,m_q)$ and $\Lambda_P(p,m_q)$
cannot be extrapolated to the chiral limit. Due to the zero modes, both
$\Lambda_S(p,m_q)$ and $\Lambda_P(p,m_q)$ diverge $\propto 1/m_q^2$. This is
an artefact of the quenched approximation. On top of that, $\Lambda_P(p,m_q)$
receives a contribution $\propto \Sigma/(m_q p^2)$. This term is due to
spontaneous chiral symmetry breaking~\cite{Pagels,QCDSF7}. We thus expect the
following dependence on the quark mass: 

\begin{eqnarray}
\Lambda_S(p,m_q) &=& \frac{C_1^S(p)}{(am_q)^2} + C_3^S(p) + C_4^S (p) \, am_q
\, , \label{lsm} \\[0.3em] 
\Lambda_P(p,m_q) &=& \frac{C_1^P(p)}{(am_q)^2} + \frac{C_2^P(p)}{am_q} +
C_3^P(p) + C_4^P(p) \, am_q \, , \label{lpm}\\[0.3em] 
\Lambda_A(p,m_q) &=& C_3^A(p) + C_4^A (p) \, am_q \, , \label{lam}
\end{eqnarray}
neglecting terms of $O(m_q^2)$. This behavior is indeed shown by the
data. In Fig.~\ref{fig:Lambda_m} we plot $\Lambda_S(p,m_q)$,
$\Lambda_P(p,m_q)$ and $\Lambda_A(p,m_q)$ for three different momenta,
together with a fit of (\ref{lsm}), (\ref{lpm}) and (\ref{lam}) to the data. 
We identify $\Lambda_S(p)$, $\Lambda_P(p)$ and $\Lambda_A(p)$ with   
$c_3^S(p)$, $c_3^P(p)$ and $c_3^A(p)$, respectively, from which we derive
\begin{equation}
Z_S(\mu) = \frac{\Lambda_A(\mu)}{\Lambda_S(\mu)} \, Z_A \, , \quad Z_P(\mu) =
\frac{\Lambda_A(\mu)}{\Lambda_P(\mu)} \, Z_A \, .
\end{equation}
We expect $Z_S(\mu) = Z_P(\mu)$ due to chiral symmetry. To test this relation,
we plot the ratio $\Lambda_S/\Lambda_P$ in Fig.~\ref{fig:Lambda_S/P}. We find
good agreement between $Z_S$ and $Z_P$ for all momenta. In the following we
shall make use of a combined fit of $\Lambda_S(p,m_q)$ and $\Lambda_P(p,m_q)$,
in which we set $C_3^S(p) = C_3^P(p)$.

We are finally interested in $Z_S$ in the $\overline{MS}$ scheme at a given
scale $\mu$. To convert our numbers from the $RI^{\prime}-MOM$ scheme, which
we were working in so far, to the $\overline{MS}$ scheme, we proceed in two
steps. In the first step we match to the scale invariant $RGI$ scheme,
\begin{equation}
Z_S^{RGI} = \Delta^{RI^{\prime}-MOM}(\mu)\, Z_S(\mu) \, ,
\end{equation}
and in the second step we evolve $Z_S^{RGI}$ to the targeted scale in the
$\overline{MS}$ scheme,
\clearpage

\begin{figure}[t]
\vspace*{0.5cm}
  \centering
  \epsfig{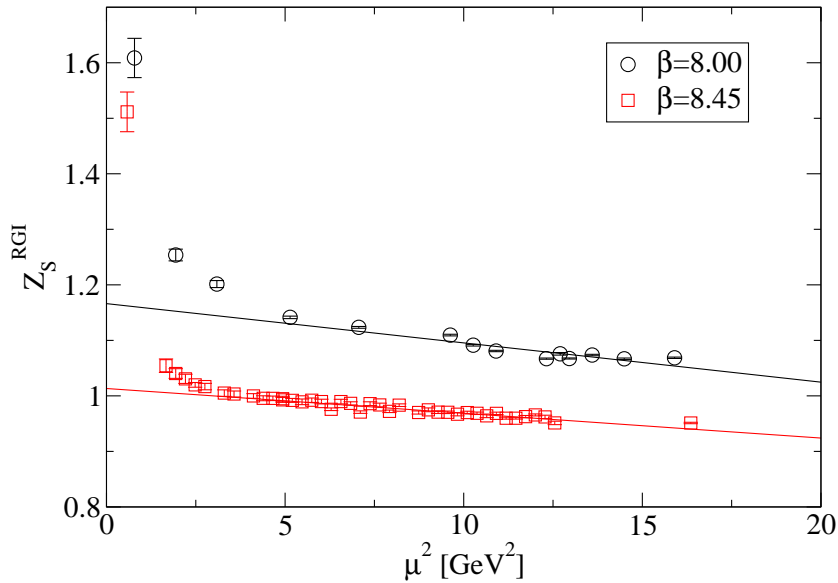}
  \caption{The scale invariant renormalization constant $Z_S^{RGI}$ on the
  $16^3 \, 32$ lattice at $\beta=8.0$ ($\Circle$) and $\beta=8.45$
  ($\square$).} 
  \label{fig:Z_S_RGI}
\end{figure}

\clearpage
\begin{equation}
Z_S^{\overline{MS}}(\mu) = \Delta^{\overline{MS}}(\mu)^{-1}\, Z_S^{RGI} \, .
\end{equation}
The matching coefficients $\Delta^{RI^{\prime}-MOM}(\mu)$ and
$\Delta^{\overline{MS}}(\mu)$ are known perturbatively to four
loops~\cite{4loop}. In Fig.~\ref{fig:Z_S_RGI} we show $Z_S^{RGI}$. The result
is not quite independent of the scale parameter $\mu$ as it should, but shows a
linear decrease in $\mu^2$ for $\mu \gtrsim 2 \, \mbox{GeV}$. We attribute
this behavior to lattice artefacts of $O(a^2\mu^2)$. Indeed, the slope of
$Z_S^{RGI}$ at our two different $\beta$ values scales like $a^2$ to a good
approximation. We thus fit the lattice result by

\begin{equation}
Z_S^{RGI, LAT} = C_0 + C_1 (a\mu)^2 
\end{equation}
and identify the physical value of $Z_S^{RGI}$ with $C_0$. This finally gives
\begin{equation}
\begin{tabular}{c|c|c}
$\beta$ & $V$ & $Z_S^{RGI}$ \\ \hline
 $8.0\phantom{0}$&  $16^3\, 32$ & 1.18(2) \\
  $8.45$&  $16^3\, 32$ & 1.02(1) \\
\end{tabular}
\label{tab:ZSRGI}
\end{equation}
The four-loop value for $\Delta_S^{\overline{MS}}(\mu)$ has been given
in~\cite{QCDSF9}. At $\mu=2 \, \mbox{GeV}$ it is
$\Delta_S^{\overline{MS}}(2 \, \mbox{GeV}) = 0.721(10)$. The error is a
reflection of the error of $\Lambda^{\overline{MS}}$. The nonperturbative 
result at $\beta=8.45$ is in good agreement with $Z_S^{FTI,RGI}$ 
from~\cite{QCDSF8}.

\section{Chiral Condensate and Quark Masses}

Having determined the renormalization constant of the scalar density, we may
now compute the renormalized chiral condensate and light and strange quark
masses.  

Let us first consider the chiral condensate $\langle \bar{\psi}\psi\rangle$. 
Strictly speaking, $\langle \bar{\psi}\psi\rangle$ is not defined in the
quenched theory due to the presence of a logarithmic singularity in the
chiral limit. Nevertheless, we may identify $ - \langle \bar{\psi}\psi\rangle$
with $\Sigma$ and assume that $\Sigma$ renormalizes like (the finite part of)
the scalar density. In the $\overline{MS}$ scheme at $\mu = 2\,\mbox{GeV}$ we
then have 
\begin{equation}
\langle \bar{\psi}\psi\rangle^{\overline{MS}}(2\,\mbox{GeV}) =
- Z_S^{\overline{MS}}(2\,\mbox{GeV})\, \Sigma \, .
\end{equation}
Taking $\Sigma$ from our second fit in (\ref{mpifit}), where we have fixed
$\delta$ to its theoretical value $0.183$, this leads to 
\begin{equation}
\begin{tabular}{c|c|c}
$\beta$ & $V$ & $\langle \bar{\psi}\psi\rangle^{\overline{MS}}(2\,\mbox{GeV})$
 \\ \hline 
 $8.0\phantom{0}$&  $16^3\, 32$ & $- \big[324(8)\phantom{0}\, \mbox{MeV}\big]^3$ \\
  $8.45$&  $24^3\, 48$ &  $- \big[296(11)\, \mbox{MeV}\big]^3$  \\
\end{tabular}
\label{tab:chicon}
\end{equation}
The lower number at the larger $\beta$ value is in reasonable agreement with
phenomenology and other quenched lattice calculations~\cite{chicon}. A better
way to determine $\Sigma$ is by means of the spectral
density~\cite{Osborn,Damgaard}, which we will address in a separate
publication~\cite{Streuer}.  

Let us now turn to the evaluation of the quark masses. We shall assume
(\ref{chipi}) with $\alpha_{\Phi}^q=0$ as the basic functional form for the
relation between the quark masses and the pseudoscalar mass:
\begin{equation}
m_{PS}^2 = X m_q + Y m_q \ln m_q + Z  m_q^2 \, .
\label{deg}
\end{equation}
For nondegenerate quark masses, $m_q^a$ and $m_q^b$, chiral perturbation theory
gives the result 
\begin{equation}
\begin{split}
\left(m_{PS}^{ab}\right)^2 &= X \left(\frac{m_q^a + m_q^b}{2}\right) + Y
  \left(\frac{m_q^a 
  + m_q^b}{2}\right) \left( 
  \frac{m_q^a \ln m_q^a - m_q^b \ln m_q^b}{m_q^a-m_q^b} -1 \right) \\&+ \,Z
\left(\frac{m_q^a + m_q^b}{2}\right)^2  
\end{split}
\label{nondeg}
\end{equation}
with no new parameter. In fact, (\ref{nondeg}) reduces exactly to (\ref{deg})
in the limit $m_q^a \rightarrow m_q^b$. We fit (\ref{deg}) to our data to
determine the coefficients $X$, $Y$ and $Z$. The light quark mass, $m_\ell =
(m_u + m_d)/2$, is then found from
\begin{equation}
m_{\pi^+}^2 =  X m_\ell + Y m_\ell \ln m_\ell + Z  m_\ell^2 \, ,
\label{light}
\end{equation}
while we compute the strange quark mass from
\begin{equation}
\begin{split}
\frac{m_{K^+}^2 + m_{K^0}^2}{2} &= X \left(\frac{m_\ell + m_s}{2}\right) + Y
  \left(\frac{m_\ell 
  + m_s}{2}\right) \left( 
  \frac{m_\ell \ln m_\ell - m_s \ln m_s}{m_\ell-m_s} -1 \right) \\&+ \;Z
\left(\frac{m_\ell + m_s}{2}\right)^2 \, . 
\end{split}
\label{strange}
\end{equation}
The result is
\begin{equation}
\begin{tabular}{c|c|c|c}
$\beta$ & $V$ & $m_\ell$ [MeV] & $m_s$ [MeV]
 \\ \hline 
 $8.0\phantom{0}$&  $16^3\, 32$ & $6.3(1)$ & $203(4)$ \\
  $8.45$&  $24^3\, 48$  & $5.3(3)$ & $160(5)$  \\
\end{tabular}
\label{tab:qm}
\end{equation}
The renormalized quark masses are given by
\begin{equation}
m_q^R = Z_m\, m_q \, ,
\end{equation}
where $Z_m = 1/Z_S$. Combining the bare quark masses in (\ref{tab:qm}) with
the results for $Z_S$ in (\ref{tab:ZSRGI}) and below, we obtain in the
$\overline{MS}$ scheme at $\mu=2\, \mbox{GeV}$
\begin{equation}
\begin{tabular}{c|c|c|c}
$\beta$ & $V$ & $m_\ell^{\overline{MS}}(2\,\mbox{GeV})$ [MeV] &
 $m_s^{\overline{MS}}(2\,\mbox{GeV})$ [MeV] 
 \\ \hline 
 $8.0\phantom{0}$&  $16^3\, 32$ & $3.8(1)$ & $124(3)$ \\
  $8.45$&  $24^3\, 48$  & $3.8(2)$ & $114(4)$  \\
\end{tabular}
\label{tab:qmmsbar}
\end{equation}
These results are in good agreement with other nonperturbative calculations of
the quark masses in the quenched approximation~\cite{Rebbi,Liu,Garden,QCDSF9}.

\section{Conclusions}

The extrapolation to the chiral limit has been a major challenge in lattice
QCD. We have shown that with using overlap fermions it is possible to
progress to small quark masses. Here we have simulated pion masses down to
$m_\pi \approx 250$ MeV on both of our lattices. We have made an attempt to
compute the low-energy constants of quenched chiral perturbation theory, with
some success. Our results turn out to be consistent with the predicted and/or
phenomenological values. To fully exploit the potential of overlap fermions 
at small quark masses, one will, however, need a statistics of several 
thousand independent gauge field configurations.

The pion mass was found to depend on the topological 
charge $|Q|$ at small quark masses. No such behavior was found for the 
pseudoscalar decay constant, but a similar effect is expected to show up in 
the chiral condensate~\cite{Osborn}.


Overlap fermions, in combination with the L\"uscher-Weisz gauge field
action, show good scaling properties already at lattice spacing $a \approx
0.15$ fm, owing to the fact that they are automatically $O(a)$ improved,
on-shell and off-shell. This helps to reduce the large numerical overhead in
the algorithm. 

The calculations performed in this paper test many of the
ingredients needed for a simulation of full QCD, and thus provide a lesson for
future applications.

\section*{Acknowledgement}
The numerical calculations have been performed on the IBM p690 at HLRN
(Berlin) and NIC (J\"ulich), as well as on the PC farm at DESY
(Zeuthen). Furthermore, we made use of the facilities on the CCHPCF at
Cambridge and of HPCx, the UK's national high performance computing service,
which is provided by EPCC at the University of Edinburgh and by CCLRC Daresbury
Laboratory, and funded by the Office of Science and Technology through EPSRC's
High End Computing Program. We thank all institutions. This work has been
supported in part by 
the EU Integrated Infrastructure Initiative Hadron Physics (I3HP) under
contract RII3-CT-2004-506078 and by the DFG under contract FOR 465
(Forschergruppe Gitter-Hadronen-Ph\"anomenologie).

%
%
%
%


\begin{thebibliography}{99}

\bibitem{Neuberger}
H. Neuberger, Phys.\ Lett.\ B417 (1998) 141; {\it ibid.} B427 (19998) 353.

\bibitem{Luscher}
M. L\"uscher, Phys.\ Lett.\ B428 (1998) 342.

\bibitem{QCDSF1}
S.~Capitani, M.~G\"ockeler, R.~Horsley, P.~E.~L.~Rakow and G.~Schierholz,
Phys.\ Lett.\ B468 (1999) 150.

\bibitem{Rebbi}
L.~Giusti, C.~Hoelbling and C.~Rebbi, Phys.\ Rev.\ D64 (2001) 114508  [Erratum
{\it ibid.}  D65 (2002) 079903];
N.~Garron, L.~Giusti, C.~Hoelbling, L.~Lellouch and C.~Rebbi,
Phys.\ Rev.\ Lett.\ 92 (2004) 042001;
F.~Berruto, N. Garron, C. Hoelbling, J. Howard, L. Lellouch, S.~Necco,
C.~Rebbi and N. Shoresh, Nucl.\ Phys. \ Proc.\ Suppl.\ 140 (2005) 264.

\bibitem{Liu}
S.~J.~Dong, F.~X.~Lee, K.~F.~Liu and J.~B.~Zhang, Phys.\ Rev.\ Lett.\ 85
(2000) 5051; 
S.~J.~Dong, T.~Draper, I.~Horvath, F.~X.~Lee, K.~F.~Liu and J.~B.~Zhang,
Phys.\ Rev.\ D65 (2002) 054507;
F.~X.~Lee, S.~J.~Dong, T.~Draper, I.~Horvath, K.~F.~Liu, N.~Mathur and
J.~B.~Zhang, Nucl.\ Phys.\ Proc.\ Suppl.\ 119 (2003) 296;
N.~Mathur, F.~X. Lee, A. Alexandru, C. Bennhold, Y.~Chen, S.J. Dong,
T. Draper, I. Horvath, K.~F. Liu, S. Tamhankar and J.~B. Zhang, Phys.\ Rev.\ D
70 (2004) 074508. 

\bibitem{QCDSF2}
D.~Galletly, M. G\"urtler, R. Horsley, B. Jo\'o, A.~D. Kennedy, H. Perlt,
B.~J. Pendleton, P.~E.~L. Rakow, G. Schierholz, A. Schiller and T. Streuer, 
Nucl.\ Phys.\ Proc.\ Suppl.\ 129 (2004) 453.

\bibitem{Bietenholz}
W.~Bietenholz, T.~Chiarappa, K.~Jansen, K.~I.~Nagai and S.~Shcheredin,
JHEP 0402 (2004) 023.
 
\bibitem{Hernandez}
P.~Hernandez, K.~Jansen and M.~L\"uscher, Nucl.\ Phys.\ B552 (1999) 363.

\bibitem{Colangelo}
G.~Colangelo, Nucl.\ Phys.\ Proc.\ Suppl.\ 140 (2005) 120.

\bibitem{QCDSF3}
D. Galletly, M. G\"urtler, R. Horsley, K. Koller, V. Linke, P.~E.~L.~Rakow,
C.~J.~Roberts, G.~Schierholz and T.~Streuer, PoS LAT2005 (2005) 363
[arXiv:hep-lat/0510050].

\bibitem{QCDSF4}
M.~G\"urtler, R. Horsley, V. Linke, H. Perlt, P. E. L. Rakow, G. Schierholz,
A. Schiller and T.~Streuer, Nucl.\ Phys.\ Proc.\ Suppl.\ 140 (2005) 707.

\bibitem{Ilgenfritz}
E.~M.~Ilgenfritz, K.~Koller, Y.~Koma, G.~Schierholz, T.~Streuer and
V.~Weinberg, arXiv:hep-lat/0512005.

\bibitem{Weinberg}
V.~Weinberg, E.~M.~Ilgenfritz, K.~Koller, Y.~Koma, G.~Schierholz and
T.~Streuer, PoS LAT2005 (2005) 171 [arXiv:hep-lat/0510056].

\bibitem{Giusti}
L.~Giusti, C.~Hoelbling, M.~L\"uscher and H.~Wittig, Comput.\ Phys.\ Commun.\
153 (2003) 31. 

\bibitem{LW}
M. L\"uscher and P. Weisz, Commun.\ Math.\ Phys.\ 97 (1985) 59.

\bibitem{Gattringer}
C. Gattringer, R. Hoffmann and S. Schaefer, Phys.\ Rev.\ D65 (2002) 094503.

\bibitem{Symanzik}
K. Symanzik, Nucl.\ Phys.\ B226 (1983) 187.

\bibitem{QCDSF5}
M.~G\"ockeler, R.~Horsley, E.-M.~Ilgenfritz, H.~Perlt, P.~Rakow, G.~Schierholz
and A.~Schiller, Phys.\ Rev.\ D53 (1996) 2317.

\bibitem{lma}
T.~DeGrand and S.~Schaefer, Comput.\ Phys.\ Commun.\ 159 (2004) 185;
L.~Giusti, P.~Hernandez, M.~Laine, P.~Weisz and H.~Wittig,
JHEP 0404 (2004) 013;
A.~O'Cais, K.~J.~Juge, M.~J.~Peardon, S.~M.~Ryan and J.~I.~Skullerud,
Nucl.\ Phys.\ Proc.\ Suppl.\ 140 (2005) 844.

\bibitem{QCDSF6}
M.~G\"ockeler, R.~Horsley, H.~Perlt, P.~Rakow, G.~Schierholz, A.~Schiller and
P.~Stephenson, Phys.\ Rev.\ D57 (1998) 5562.

\bibitem{Pallante}
G.~Colangelo and E.~Pallante, Nucl.\ Phys.\ B520 (1998) 433.

\bibitem{Bernard}
C.~W.~Bernard and M.~F.~L.~Golterman, Phys.\ Rev.\ D46 (1992) 853.

\bibitem{Sharpe}
S.~R.~Sharpe, Phys.\ Rev.\ D46 (1992) 3146.

\bibitem{Gasser}
J.~Gasser and H.~Leutwyler, Nucl.\ Phys.\ B250 (1985) 465.


\bibitem{Bijnens}
J.~Bijnens, AIP Conf.\ Proc.\  768 (2005) 153  [arXiv:hep-ph/0409068].

\bibitem{Sommer}
J.~Heitger, R.~Sommer and H.~Wittig, Nucl.\ Phys.\ B588 (2000) 377.

\bibitem{WV} E.~Witten, Nucl.\ Phys.\ B156 (1979) 269; G.~Veneziano, Nucl.\
Phys.\ B159 (1979) 213. 

\bibitem{Brower}
R.~Brower, S.~Chandrasekharan, J.~W.~Negele and U.~J.~Wiese,
Phys.\ Lett.\ B560 (2003) 64.

\bibitem{Booth}
M.~Booth, G.~Chiladze and A.~F.~Falk, Phys.\ Rev.\ D55 (1997) 3092.

\bibitem{Labrenz}
J.~N.~Labrenz and S.~R.~Sharpe, Phys.\ Rev.\ D54 (1996) 4595.

\bibitem{Martinelli}
G.~Martinelli, C.~Pittori, C.~T.~Sachrajda, M.~Testa and A.~Vladikas,
Nucl.\ Phys.\ B445 (1995) 81.

\bibitem{QCDSF7}
M.~G\"ockeler, R. Horsley, H. Oelrich, H. Perlt, D. Petters, P.~E.~L. Rakow,
A. Sch\"afer, G.~Schierholz and A. Schiller, Nucl.\ Phys.\ B544 (1999) 699.

\bibitem{QCDSF8}
R.~Horsley, H.~Perlt, P.~E.~L.~Rakow, G.~Schierholz and A.~Schiller, Nucl.\
Phys.\ B693 (2004) 3  [Erratum {\it ibid.\ } B713 (2005) 601].

\bibitem{Pagels}
H.~Pagels, Phys.\ Rev.\ D19 (1979) 3080.

\bibitem{4loop}
K.~G. Chetyrkin, Phys.\ Lett.\ B404 (1997) 161; J.~A.~M Vermaseren,
S.~A. Larin and T.~van Ritbergen, Phys.\ Lett.\ B405 (1997) 327;
K.~G. Chetyrkin and A. R\'etey, Nucl.\ Phys.\ B583 (2000) 3; see also
M. G\"ockeler, R. Horsley, A.~C. Irving, D. Pleiter, P.~E.~L. Rakow,
G.~Schierholz, H. St\"uben and J.~M.~Zanotti, arXiv:hep-lat/0601004.

\bibitem{QCDSF9}
 M. G\"ockeler, R. Horsley, H. Oelrich, D. Petters, D. Pleiter,
 P.~E.~L. Rakow, G.~Schierholz, P.~Stephenson  M.~G\"ockeler, Phys.\ Rev.\ D62
 (2000) 054504.

\bibitem{Osborn}
J.~C.~Osborn, D.~Toublan and J.~J.~M.~Verbaarschot, Nucl.\ Phys.\ B540 (1999)
317.

\bibitem{Damgaard}
P.~H.~Damgaard, Nucl.\ Phys.\ B608 (2001) 162.

\bibitem{chicon}
L.~Giusti, F.~Rapuano, M.~Talevi and A.~Vladikas, 
Nucl.\ Phys.\ B538 (1999) 249;
T.~A.~DeGrand, Phys.\ Rev.\ D64 (2001) 117501;
P.~Hernandez, K.~Jansen, L.~Lellouch and H.~Wittig, 
JHEP 0107 (2001) 018; 
L.~Giusti, C.~Hoelbling and C.~Rebbi, Phys.\ Rev.\ D64 (2001) 114508
[Erratum-ibid.\ D65 (2002) 079903];
T.~Blum, P.~Chen, N.~Christ, C.~Cristian, C.~Dawson, G.~Fleming, A.~Kaehler, 
X.~Liao, G.~Liu, C.~Malureanu, R.~Mawhinney, S.~Ohta, G.~Siegert, A.~Soni, 
C.~Sui, P.~Vranas, M.~Wingate, L.~Wu and Y.~Zhestkov, 
Phys.\ Rev.\ D69 (2004) 074502;
D.~Becirevic and V.~Lubicz, Phys.\ Lett.\ B600 (2004) 83;
V.~Gimenez, V.~Lubicz, F.~Mescia, V.~Porretti and J.~Reyes, 
Eur.\ Phys.\ J.\ C41 (2005) 535.

\bibitem{Streuer}
M.~G\"urtler et al., in preparation.

\bibitem{Garden}
J.~Garden, J.~Heitger, R.~Sommer and H.~Wittig, Nucl.\ Phys.\ B571 (2000) 237.

\end{thebibliography}
\end{document}